\documentclass[fleqn,usenatbib]{mnras}

\usepackage{newtxtext,newtxmath}

\usepackage[T1]{fontenc}

\DeclareRobustCommand{\VAN}[3]{#2}
\let\VANthebibliography\thebibliography
\def\thebibliography{\DeclareRobustCommand{\VAN}[3]{##3}\VANthebibliography}

\usepackage{graphicx}	
\usepackage{amsmath}	
\usepackage{rotating}
\usepackage{wasysym}    

\usepackage{orcidlink}
\usepackage{xspace}
\usepackage{ulem}
\makeatletter 
  \patchcmd{\NAT@citex}
    {\@citea\NAT@hyper@{%
      \NAT@nmfmt{\NAT@nm}%
      \hyper@natlinkbreak{\NAT@aysep\NAT@spacechar}{\@citeb\@extra@b@citeb}%
      \NAT@date}}
    {\@citea\NAT@nmfmt{\NAT@nm}%
    \NAT@aysep\NAT@spacechar\NAT@hyper@{\NAT@date}}{}{}

  \patchcmd{\NAT@citex}
    {\@citea\NAT@hyper@{%
      \NAT@nmfmt{\NAT@nm}%
      \hyper@natlinkbreak{\NAT@spacechar\NAT@@open\if*#1*\else#1\NAT@spacechar\fi}%
        {\@citeb\@extra@b@citeb}%
      \NAT@date}}
    {\@citea\NAT@nmfmt{\NAT@nm}%
    \NAT@spacechar\NAT@@open\if*#1*\else#1\NAT@spacechar\fi\NAT@hyper@{\NAT@date}}
    {}{}
\makeatother

\newcommand{\HM}{\ion{H}{$_2$}\xspace}
\newcommand{\HI}{\ion{H}{I}\xspace}
\newcommand{\HII}{\ion{H}{II}\xspace}
\newcommand{\HeI}{\ion{He}{I}\xspace}
\newcommand{\HeII}{\ion{He}{II}\xspace}
\newcommand{\HeIII}{\ion{He}{III}\xspace}

\newcommand{\CI}{\ion{C}{I}\xspace}
\newcommand{\CII}{\ion{C}{II}\xspace}
\newcommand{\CIII}{\ion{C}{III}\xspace}

\newcommand{\CVI}{\ion{C}{VI}\xspace}

\newcommand{\NI}{\ion{N}{I}\xspace}
\newcommand{\NII}{\ion{N}{II}\xspace}

\newcommand{\NVII}{\ion{N}{VII}\xspace}

\newcommand{\OI}{\ion{O}{I}\xspace}
\newcommand{\OII}{\ion{O}{II}\xspace}
\newcommand{\OIII}{\ion{O}{III}\xspace}

\newcommand{\OVIII}{\ion{O}{VIII}\xspace}

\newcommand{\SII}{\ion{S}{II}\xspace}

\newcommand{\SIV}{\ion{S}{IV}\xspace}

\newcommand{\FeII}{\ion{Fe}{II}\xspace}

\newcommand{\thesan}{\textsc{thesan}\xspace}
\newcommand{\thzoom}{\mbox{\textsc{thesan-zoom}}\xspace}
\newcommand{\colt}{\textsc{colt}\xspace}
\newcommand{\jwst}{\textit{JWST}\xspace}

\title[Nebular emission with COLT]{
Modelling the nebular emission of galaxies across cosmic time with COLT
}

\author[W. McClymont, Smith \& Tacchella]{%
William McClymont$\orcidlink{0009-0009-5565-3790}$,$^{1,2}$\thanks{E-mail: \href{mailto:wjm50@cam.ac.uk}{wjm50@cam.ac.uk} (WM)}
Aaron Smith$\orcidlink{0000-0002-2838-9033}$$^{3}$\thanks{E-mail: \href{mailto:asmith@utdallas.edu}{asmith@utdallas.edu} (AS)} and
Sandro Tacchella$\orcidlink{0000-0002-8224-4505}$$^{1,2}$
\\
\\
$^{1}$Kavli Institute for Cosmology, University of Cambridge, Madingley Road, Cambridge CB3 0HA, UK\\
$^{2}$Cavendish Laboratory, University of Cambridge, 19 JJ Thomson Avenue, Cambridge CB3 0HE, UK\\
$^3$ Department of Physics, The University of Texas at Dallas, Richardson, TX 75080, USA \\
}

\date{Accepted XXX. Received YYY; in original form ZZZ}

\pubyear{\the\year{}}

\begin{document}
\label{firstpage}
\pagerange{\pageref{firstpage}--\pageref{lastpage}}
\maketitle

\begin{abstract}
Extragalactic nebular emission has long been a workhorse probe of the processes driving galaxy evolution, but the richness of \textit{JWST} spectroscopy has shifted the bottleneck from data acquisition to physical interpretation and modelling. In this context, we present a major update to the Monte Carlo radiative transfer code \textsc{colt} to facilitate self-consistent modelling of nebular line and continuum emission from simulated galaxies. We introduce a new thermal equilibrium solver that iteratively couples to the existing ionization solver and radiation field to compute effective gas temperatures by accurately balancing photoionization heating, radiative and dielectronic recombination, collisional ionization, charge exchange, metal and primordial line cooling, free-free emission, and Compton scattering. To prevent over-cooling where non-equilibrium hydrodynamics dominate, we introduce a Courant-limited cooling prescription tied to each cell's sound-crossing time, preserving temperatures in the diffuse halo while allowing physically motivated cooling in the interstellar medium (ISM). Applied to an isolated local galaxy simulation, the equilibrium solver reshapes the ISM phase space by reducing spuriously excessive lukewarm ($T=10^3-10^4$\,K) gas and better resolving warm ionized and cold neutral phases, while leaving the CGM largely intact. We further implement a level population solver based on modern atomic data, enabling accurate cooling and emissivities for a large library of UV to infrared metal lines, together with newly implemented primordial nebular continuum emission from free-free, free-bound, and two-photon processes. Finally, by applying \textsc{colt} to the high-redshift \textsc{thesan-zoom} simulations, we reproduce observed emission-line ratios, establishing \textsc{colt} as a robust framework for forward modelling nebular emission across cosmic time.
\end{abstract}




\section{Introduction}
\label{sec:Introduction}

The interstellar medium (ISM) is the engine that powers the build-up of galaxies across cosmic time. Its state determines whether a galaxy is star-forming or quiescent, whether its black hole grows or stagnates, and whether its radiation escapes or is smothered. In the local Universe, large integrated \citep[e.g.][]{Kauffmann:2003aa,Brinchmann:2004aa} and spatially resolved \citep[e.g.][]{Bundy:2015aa,Emsellem:2022aa} spectroscopic surveys have allowed us to peer inside the ISM with nebular emission lines \citep{Kewley:2019aa}, revealing fundamental properties such as density \citep[e.g.][]{Kewley:2019ab}, metallicity \citep[e.g.][]{Maiolino:2019aa}, and the shape of the permeating ionizing radiation field \citep[e.g.][]{Baldwin:1981aa,Belfiore:2016aa}.

Extending this work to study the ISM of galaxies in the early Universe is a frontier of modern astrophysics. Rapid progress in this regime has been made in recent years due to the extreme infrared sensitivity of \jwst, which has allowed us to study the rest-frame UV and optical emission line characteristics of galaxies deep into the Epoch of Reionization \citep[EoR;][]{Stark:2025aa}. This wealth of high-redshift emission line data has unveiled the extreme internal lives of galaxies in the early universe, with high ionization parameters \citep{Reddy:2023aa}, hard ionizing spectra \citep{Topping:2024aa}, and extremely low metallicity \citep{Curti:2023aa,Curti:2024ab}.

While \jwst has unquestionably driven a revolution in the field, in many ways it has raised more questions than it has answered. For example, the origin of the high ionization parameters and hard spectra is strongly debated, with both active galactic nuclei \citep[AGN;][]{Maiolino:2024ac,Chisholm:2024aa,Scholtz:2025aa} and various extreme modes of star formation being proposed as solutions \citep{Bunker:2023aa,Curti:2025aa}. There are also indications that the duty cycle of star formation in the early universe is incredibly short, on the scale of tens of Myr \citep{Tacchella:2023aa,Looser:2024aa,Looser:2025aa,McClymont:2025aa,Simmonds:2025aa,Witten:2025aa}, however, the magnitude and origin of this behaviour is still debated. \jwst has observed some particularly strange spectral features, such as non-Case B emission \citep{Scarlata:2024aa,Yanagisawa:2024aa,McClymont:2025ad}, and potentially extreme Lyman damping \citep{Li:2024aa,Tacchella:2025aa,Witten:2025ab} or nebular continuum emission \citep{Cameron:2024aa,Katz:2024aa}, which may be due to vast reservoirs of neutral gas in the ISM or a top-heavy stellar initial mass function (IMF), respectively. \jwst has also found a lack of widespread rotationally-supported disks \citep{Danhaive:2025aa}, in apparent tension with results from the \textit{Atacama Large Millimeter Array} \citep[\textit{ALMA}; e.g.,][]{Lelli:2021aa,Pope:2023aa,Rowland:2024ab}. We have also observed peculiar chemical enrichment patterns, such as super-solar nitrogen-to-oxygen ratios (N/O) in low metallicity galaxies \citep[e.g.][]{Bunker:2023aa,Cameron:2023ab,Isobe:2023aa,Ji:2024aa,Schaerer:2024aa}. A variety of explanations have been put forward to explain these abundances, including fast-rotating massive stars \citep{Vink:2023aa,Nandal:2024ab,Tsiatsiou:2024aa}, very or extremely massive stars \citep[$M_\star>100~\text{M}_\odot$;][]{Charbonnel:2023aa,Nagele:2023aa,Nandal:2024aa,Nandal:2025aa,Gieles:2025aa}, Wolf-Rayet stars \citep{Kobayashi:2024aa,Watanabe:2024aa}, tidal disruption events \citep{Cameron:2023ab}, AGB star enrichment combined with differential outflows \citep{Rizzuti:2025aa} or pristine gas inflows \citep{DAntona:2023aa}, and N/O enhancement as a natural result of bursty star formation \citep{McClymont:2025ae}, however, a consensus has yet to be reached.

While these unresolved challenges have only recently reared their heads, they drive at fundamental questions which have been at the heart of the field for decades; how do black holes and their host galaxies grow? How is star formation regulated in galaxies? How did the Universe become enriched with metals? The extreme nature of galaxies in the early Universe has made it clear that there is still much work to be done before we have definitive and comprehensive answers to these questions.

All of the challenges outlined above have been identified through observations of nebular emission. In fact, we now have such a variety of puzzling observations that progress is increasingly limited by modelling rather than by observational data. Typically, emission lines are interpreted using photoionization models, such as \textsc{Cloudy} \citep{Ferland:2017aa,Gunasekera:2025aa} and \textsc{mappings} \citep{Allen:2008aa,Sutherland:2018aa}, which simulate the physical conditions in ionized gas regions. While these codes are invaluable for detailed studies of localized \ion{H}{II} regions, they have serious limitations when applied on a galaxy-wide scale. These models often assume uniform conditions, leading to discrepancies in reproducing observed emission line ratios \citep{Cameron:2023ac,Choustikov:2025ab}. They also cannot self-consistently model emission from gas which is spatially disconnected from its ionizing source, such as the diffuse ionized gas \citep[DIG; e.g.,][]{Wood:2004aa,Wood:2010aa,Kado-Fong:2020aa,Belfiore:2022aa,Tacchella:2022aa,McClymont:2024aa,McCallum:2024aa,McCallum:2024ab}. 

To overcome these limitations, modelling emission lines on a galaxy-wide scale using high-resolution hydrodynamic simulations has become a promising approach. Such simulations allow for a more nuanced treatment of the ISM, incorporating variations in physical conditions across different regions of a galaxy. While this approach has a great number of benefits, modelling the emission from simulated galaxies accurately poses a significant challenge, requiring both the calculation of the ionization states in a galaxy and the observed emission arising from these ions.

On-the-fly calculation of ionization states, including metals, in radiation-hydrodynamic (RHD)\footnote{It is possible to include non-equilibrium or quasi-non-equillibrium metal cooling without full RHD \citep[e.g.,][]{Glover:2007aa,Richings:2022aa,Thompson:2024aa,Ploeckinger:2025aa,Schaye:2025aa}, even restricted to the most important cooling channels like \CII\ \citep[e.g.,][]{Deng:2024ab}.} simulations is perhaps the obvious method, and presents other advantages such as the non-equilibrium cooling due to metal lines rather than using pre-tabulated rates assuming photoionization equilibrium \citep{Katz:2022ad,Katz:2022ac,Katz:2024aa,Katz:2025ab,Choustikov:2025ab}. This method is certainly a promising approach, however, it is computationally expensive, which limits its application. Additionally, on-the-fly solvers are limited to a few energy bins, preventing them from accurately capturing the shape of the radiation field \citep[e.g.,][]{Rosdahl:2013aa,Kannan:2022aa,Kannan:2025aa,Wadsley:2024aa}. While this is not catastrophic for capturing the ionization of H and He, ions such as \OIII have ionization potentials that generally fall within radiation bins, and therefore their large-scale ionization cannot be faithfully reproduced \citep{McClymont:2024aa}. In addition to the limited spectral resolution of on-the-fly solvers, the typically employed M1 closure approximation \citep{Levermore:1984aa} means that the spatial distribution of radiation is necessarily imperfectly modeled \citep{Smith:2022aa}. Finally, the on-the-fly solver is particularly vulnerable to incorrectly modelling the ionization in Str\"omgren spheres when the spatial and temporal resolution of the multiphase gas thermochemistry is not sufficient \citep{Smith:2022aa,Deng:2024aa,Ejdetjarn:2024aa}. To alleviate these issues, solutions such as applying a minimum temperature floor when calculating line emission from snapshots can be used, although this reduces the self-consistency of the method \citep{Katz:2022ad}. Other strategies are to decouple the radiation treatment from other physics, or even stochastically photoheat gas to $10^4$\,K to enforce the expected behaviour \citep{Hopkins:2019aa,Hopkins:2020aa,Marinacci:2019aa,Smith:2021aa}.

Post-processing offers significant advantages, allowing for much more expensive calculations than on-the-fly methods in order to confront many of the issues outlined above. One simpler post-processing method involves substituting one-dimensional, isolated models of \ion{H}{II} region emission in place of young stellar particles \citep[e.g.,][]{Hirschmann:2017aa,Hirschmann:2023aa,Hirschmann:2023ab,Pellegrini:2020aa,Pellegrini:2020ab,Vogelsberger:2020ab,Kapoor:2023aa,Lovell:2024aa,Lovell:2025aa}. This method has the benefit that it can be applied fairly safely even to simulations that do not have a resolved ISM. However, this method still relies on treating line emission as originating purely from discrete, isolated, and spherically symmetric \ion{H}{II} regions. Additionally, the impact of leaking radiation and DIG, both of which can strongly affect emission line ratios \citep[e.g.,][]{Zackrisson:2013aa,Zhang:2017aa,Belfiore:2022aa,McClymont:2024aa,McClymont:2025ad}, is not self-consistently treated, if at all. This gas is also important for modelling the morphology of line emission \citep{McClymont:2025ab}, which is particularly important given recent \jwst results showing that the stellar and nebular emission of high-redshift galaxies have distinct morphology \citep[][Villanueva et al. in preparation]{Danhaive:2025ab}.

More advanced post-processing methods run calculations directly on gas cells from a hydrodynamic galaxy simulation. This method is much more accurate but requires high-resolution simulations that resolve the ISM and sophisticated post-processing codes for detailed radiative transfer calculations. For example, the \textsc{sphinx} simulation \citep{Rosdahl:2018aa} approach uses the on-the-fly calculated radiation field as an input for \textsc{Cloudy}, which is used to calculate the ionization states and emission from cells \citep{Katz:2023ab}. A similar approach is used in the \textsc{serra} simulations \citep{Pallottini:2019aa,Pallottini:2022aa}. This approach requires assumptions about the ionizing spectral shape within each bin, which may have a significant impact on the relative abundances of ionization states, and is of course still limited by the fidelity of the on-the-fly RT. Additionally, where cells are identified as having unresolved Str\"omgren spheres, \textsc{sphinx} substitutes \textsc{Cloudy} models of full \HII regions. While this does help to alleviate some issues with the on-the-fly solver, it reduces the self-consistency of the method and introduces the issues typical of the less advanced post-processing methods discussed above. This subgrid solution is particularly necessary given the relatively low stellar particle masses compared to the gas resolution typically used in adaptive mesh refinement simulations such as \textsc{sphinx}. 

The Cosmic Ly$\alpha$ Transfer code \citep[\colt;][]{Smith:2015aa,Smith:2019aa,Smith:2022aa} is a Monte Carlo radiative transfer (MCRT) code which represents a complementary first-principles only strategy for the postprocessing of galaxy simulations, where ionization states are self-consistently recalculated assuming photoionization equilibrium. Other MCRT postprocessing codes have been applied to study galaxy evolution, such as \textsc{Sunrise} \citep{Jonsson:2006aa}, \textsc{Hyperion} \citep{Robitaille:2011aa}, \textsc{radmc-3d} \citep{Dullemond:2012aa}, \textsc{skirt} \citep{Camps:2015aa,Camps:2020aa}, \textsc{ART$^2$} \citep{Li:2020aa}, \textsc{rascas} \citep{Michel-Dansac:2020aa}, \textsc{Powderday} \citep{Narayanan:2021aa}, and \textsc{thor} \citep{Byrohl:2025aa}, however (the publicly available versions of) these codes are typically only used to calculate observable emission, rather than the ionization state of the gas. Advanced ionization treatments are included in some MCRT codes designed for simulating subgalactic scales, such as \textsc{mocassin} \citep{Ercolano:2003aa}, \textsc{sedona} \citep{Kasen:2006aa}, \textsc{torus} \citep{Harries:2019aa}, \textsc{CMacIonize} \citep{Vandenbroucke:2018ab}, and \textsc{sirocco} \citep{Matthews:2025aa}, or intergalactic scales, such as \textsc{crash} \citep{Maselli:2003aa,Ma:2022aa}. The \textsc{colt} approach involves recalculating the radiation field with a more accurate MCRT solver compared to the M1 method commonly used in on-the-fly solvers. The MCRT method is also used to calculate the observed emission, allowing for the self-consistent treatment of scattering and absorption by a heterogeneous dust distribution. Applications of \textsc{colt} at low redshifts have demonstrated its capability to produce realistic hydrogen emission line properties \citep{Smith:2022aa,Tacchella:2022aa} and study absorption lines in outflows \citep{Carr:2025aa}.

One particular issue with advanced post-processing methods is that they rely on the temperature calculated on the fly. This temperature can be inaccurate due to various factors, such as the inaccuracy of the radiation field or insufficient spatial or temporal resolution. Numerically transient, unphysical temperatures can be catastrophic for the treatment of collisionally excited metal lines, which have emissivities that are exponentially dependent on temperature. This sensitivity affected our previous work using \colt to study metal lines in a Milky Way-like galaxy \citep{McClymont:2024aa}. In that work, we were limited to studying the emission from diffuse gas, which is not prone to the unresolved Str\"omgren sphere issue. In this work, we overcome this restriction by implementing a new thermal equilibrium solver in \colt, where we allow the relaxation to an effective equilibrium temperature in post-processing. The goal of this approach is to remove temperature artifacts, and limitations are applied to ensure that gas with strong non-equilibrium effects is not over-cooled.

In the aforementioned application in \citet{McClymont:2024aa}, the metal line emissivities were based on widely-used functional forms rather than employing a full level population solver. This limits both the accuracy of the rates and the number of lines that can be calculated as they require prior calculations in the literature. In this work, we generalize the implementation with a new level population solver for metal ions. In addition, we have implemented the capability for \colt to model nebular continuum emission due to free--free, free--bound, and two-photon processes. This is now particularly relevant due to the new abundance of spectra showing Balmer jumps \citep{Roberts-Borsani:2024aa} and purportedly nebular-dominated galaxies \citep{Cameron:2024aa,Katz:2024aa}. In addition to these headline features, we have implemented a number of other improvements to the overall codebase related to continuum and line emission, such as the ability to generate full spectra.

Overall, this paper represents a major extension of the capabilities of \colt. Our goal is to produce accurate, high-fidelity models of nebular emission from galaxies across cosmic time and confront the challenges by \jwst with apples-to-apples comparisons. In this work, we will demonstrate how the new capabilities of \colt mean it is well suited to achieve this goal by showcasing its capability across a variety of proof-of-concept applications, from simple Str\"omgren spheres, to a Large Magellanic Cloud-like (LMC-like) galaxy simulation, to the \thzoom simulation suite of high-redshift galaxies. 

The paper is organised as follows. In Section~\ref{sec:COLT}, we give an overview of \colt, including the newly implemented features such as the thermal equilibrium solver, level population solver, and nebular continuum emission. In Section~\ref{sec:Model validation}, we present collisional ionization equilibrium and Str\"omgren sphere tests of \colt. In Section~\ref{sec:LMC Simulation}, we demonstrate the application of the newly implemented features to an LMC-like simulation. In Section~\ref{sec:Nebular emission of high-redshift galaxies}, we apply \colt to a limited sample of galaxies from the \thzoom simulation in order to demonstrate the code's ability to recreate observations, and we discuss the physical origin of the line ratios observed in high-redshift galaxies. We conclude in Section~\ref{sec:Conclusions}, where we summarise our main results.

\section{COLT}
\label{sec:COLT}

\begin{figure*}
    \centering
	\includegraphics[width=\textwidth]{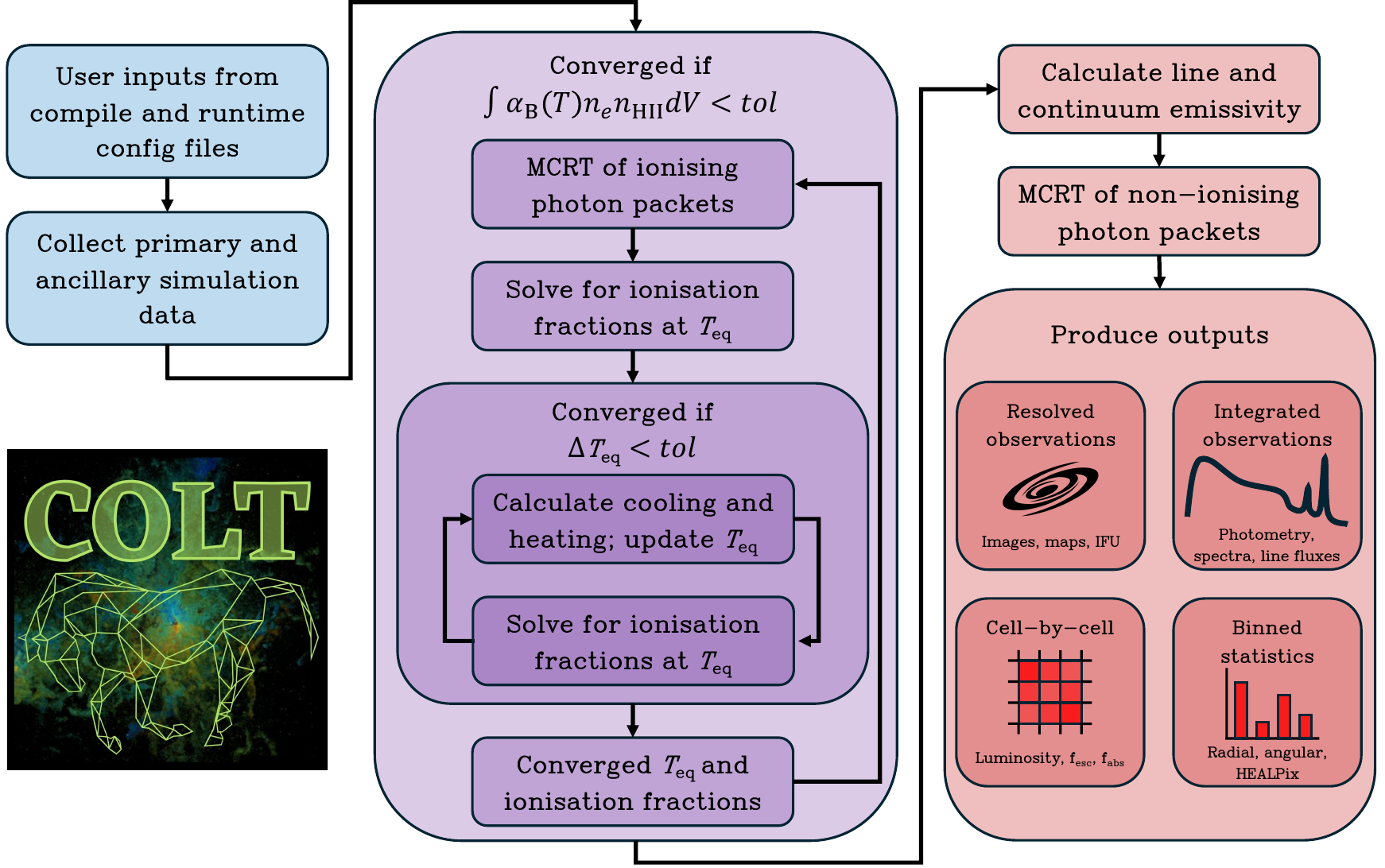}
    \caption{A schematic overview of the \colt post-processing workflow, including new features introduced in this work. User-defined inputs and simulation data set the simulation parameters, including gas and dust properties and source SEDs. The core code loop begins with ionizing radiation sampled from sources being transported through the gas, which accumulates photoionization and photoionization heating rates. These rates, along with rates for collisional ionization, radiative and dielectronic recombination, and charge exchange, are then plugged into the ionization solver to find the equilibrium ionization states. These initial states are then fed into the thermal equilibrium solver, which uses a root-finding algorithm (see text for details) to balance heating and cooling due to photoionization heating, radiative and dielectronic recombination, collisional ionization, charge exchange, metal and primordial line cooling, free-free emission, and Compton scattering. The updated, converged temperature and ionization states alter the transport of ionizing radiation, and so we iterate the core code loop until the change in the total number of hydrogen recombinations ($\int\alpha_\text{B} n_e n_\mathrm{HII}\, \text{d}V$) is less than a user-defined tolerance. After this joint convergence of the radiation field, ionization, and temperature, line emissivities are evaluated with the level-population solver and, where requested, non-ionizing stellar and nebular continuum is transported to produce mock observables (e.g., maps, IFS cubes, integrated spectra, profiles, sizes) alongside cell-level (ion fractions, temperatures, heating/cooling rates, escape/absorption fractions), star-level (escape/absorption/integrations), photon-level (paths/scattering), and binned (radial, angular, HEALPix) statistics.}
    \label{fig:colt_cartoon}
\end{figure*}

The Cosmic Lyman-alpha Transfer (\colt)\footnote{For public code access and documentation see \href{https://colt.readthedocs.io}{\texttt{colt.readthedocs.io}}.} \citep{Smith:2015aa,Smith:2019aa,Smith:2022aa} is an MCRT code that can model the absorption, emission, and scattering of arbitrary radiation sources and media, including an iterative equilibrium solver for the ionization states and temperature of the gas. \colt can also be used to produce mock observations of stellar continuum, nebular continuum, and nebular line emission, as well as other data analysis utilities. The code is designed to be applied in post-processing to (radiation-)hydrodynamic simulations with adaptive and unstructured meshes, though it can also be used on simpler geometries, such as plane parallel slabs, spherical geometry, and 3D Cartesian grids. \colt employs a efficient hybrid MPI + OpenMP parallelization strategy within a flexible C++ framework.

We here give an overview of \colt. The ionization solver and radiative transfer algorithms are largely unchanged from previous works, although there are some useful improvements. The focus of this work is the implementation of a new thermal equilibrium solver and new radiation sources (nebular continuum, metal line emission).

\subsection{Sources and spectra}
\label{sec:Sources and spectra}

\colt allows for a wide range of sources, including simplified test spectra such as a blackbody or arbitrary tabulation from geometric points, surfaces, and volumes. However, for applications to simulations, sources are generally characterized by spectral energy distributions (SEDs) represented by IMF-averaged stellar populations. In addition to stellar sources, \colt can also include cell-based emission, such as nebular continuum. However, our current implementation uses Case B rates which assume on-the-spot absorption of the ionizing component, and so it is not appropriate to incorporate this emission, although we plan to relax this assumption in the future.

In simulations, there are usually a large number of sources with a wide range of luminosities (depending on stellar age). To ensure the emission is well-sampled with a sufficient number of directions from less luminous sources, we allow the option to apply a luminosity boosting technique that biases the MCRT source probability distribution and corresponding weight assignments by a power law proportional to $\propto L^\beta$ for sampling photon packets, with a typical value of $\beta = 3/4$. This is done on a per source and per frequency bin basis, allowing for a more balanced and robust sampling across the entire spectrum of sources and frequencies.

The shape of the radiation field has a major impact on the relative abundance of different ionization states and on the photoionization heating of the gas. It is usual for on-the-fly radiation transport solvers to use very coarse spectral resolution due to computational limitations, however, \colt allows for arbitrary resolution of the ionizing radiation field in postprocessing. This allows for us to properly capture the effects of energy-dependent cross sections and radiation hardening on the shape of the radiation field. We firstly include the ionization potential energy of each tracked ion as an energy bin edge. We further subdivide bins by enforcing a minimum logarithmic bin width, e.g. less than 0.05\,dex. We apply a frequency biasing scheme of $\nu^\alpha$, typically with $\alpha = 2$, when sampling the bin to better probe the high-energy end of the SED. We also ensure a minimum fraction of photon packets, e.g. 80 per cent, in terms of the overall cumulative frequency distribution are hydrogen ionizing with energies above 13.6\,eV.
In addition to ionizing radiation followed with MCRT, \colt can optionally include an ultra-violet background (UVB). The UVB of \citet{Faucher-Giguere:2009aa} is currently included in the code, and can be specified with or without self-shielding following \citet{Rahmati:2013aa}. Other options including compatibility for each metal species are available by reading user-provided SEDs.

\subsection{Ionization and recombination}
\label{sec:Ionization and recombination}

The ionization solver takes into account a wide range of physical processes across a large chemical network. The default chemical network includes H, He, C, N, O, Ne, Mg, Si, S, and Fe, although it is possible to only use a subset of these based on ionic species ranges or ionization thresholds. Photoionization is accounted for via the transport of ionizing radiation through the gas (see Section~\ref{sec:Radiative transfer}). Photoionization cross sections are taken from \citet{Verner:1996aa}.

 For H and He, Case B recombination rates are used from \citet{Hui:1997aa}. Metal recombination rates are taken from \citet{Badnell:2006aa}. Dielectronic recombination is included using rates from \citep{Badnell:2006aa}. Both primordial and metal collisional ionization rates are taken from \citet{Voronov:1997aa}. Charge exchange reactions are taken into account between metal ions and \HI, \HII, \HeI, \HeII, and \HeIII, as well as been between \HI and He, with rates are taken from \citet{Kingdon:1996aa}. This process can be particularly important for ions which have similar ionization potentials, such as \HI and \OI. 

\subsection{Dust}
\label{sec:Dust}

\colt allows for a flexible approach to dust depending on the simulation to which it is being applied. For simulations with on-the-fly dust modelling, the dust distribution can be taken directly from the simulation. Otherwise, the dust mass in a cell can be defined with a user-defined dust-to-metal or dust-to-gas ratio. The dust opacity, albedo, and anisotropic scattering parameters can be defined by a particular dust model, such as from the Milky Way dust model of \citet{Weingartner:2001aa} or user-provided tables. Alternatively, different parameterizations can be used self-consistently for different multispecies dust components, such as carbonaceous and silicate grains and cyclic aromatic hydrocarbons (PAHs). This is particularly appropriate for simulations that include on-the-fly dust modelling with different components, such as \thzoom.

\subsection{Radiative transfer}
\label{sec:Radiative transfer}

When an ionizing photon packet is sampled from a star in a simulation, it is launched into the ISM with an isotropic randomly assigned direction, starting from the position of the star. The gas cells within the simulation are represented by a Voronoi tessellation (or another geometry), and we employ ray-tracing techniques to accurately propagate the photon through the ISM, as introduced to \textsc{colt} by \citet{Smith:2017aa}. We thus maintain a faithful representation of the \textsc{arepo} geometry and simulated gaseous and dusty components of the ISM.

Each photon packet in the simulation is assigned an initial weight. As it traverses through the ISM, this photon packet can undergo various interactions, including attenuation by dust, scattering by dust, or photoionization of atoms and ions in the gas. Absorption processes are modelled continuously, with the weight of the photon packet decreasing exponentially by $e^{-\tau_\text{a}}$ for each distance segment $\Delta\ell$ it travels. Here, $\tau_\text{a}$ is the total pure absorption optical depth along that distance in a particular energy bin. Specifically, this is given as the sum of the absorption coefficients for dust and gas, i.e. $k_\text{a} \equiv \text{d}\tau_\text{a}/\text{d}\ell = (1 - A) \kappa_\text{d} \mathcal{D} \rho + \sum_x n_x \sigma_x$, where $A$ denotes the scattering albedo, $\kappa_\text{d}$ the dust opacity, $\mathcal{D}$ the dust-to-gas ratio, $\rho$ the gas density, $n_x$ the number densities of each ionic species $x$, and $\sigma_x$ the photoionization cross-sections of the ion in the given band \citep[for details see][]{Smith:2022aa}.

The remaining MCRT procedures within the simulation are fairly standard. Dust scattering is accounted for by determining a scattering distance using the exponential distribution, with a scattering coefficient given by $k_\text{s} \equiv \text{d}\tau_\text{s}/\text{d}\ell = A \kappa_\text{d} \mathcal{D} \rho$ for the relevant energy band.
Photon trajectories are terminated when its weight becomes negligible, which by default is set to be when it falls below $10^{-14}$ times the total ionizing photon emission rate from stars.
In addition to the stellar photoionization already described, the ionization equilibrium solver also accounts for recombinations, collisional ionizations, and charge exchange reactions. Finally, as the transport opacities depend on the ionization states, \textsc{colt} employs iteration between MCRT (e.g.\ with $10^8$ photon packets) and abundance updates until the global number of recombinations is converged to within e.g.\ $0.1$ per cent.

\subsection{Cooling and heating processes}
\label{sec:Cooling and heating processes}

\subsubsection{Ionization and recombination}
\label{sec:Ionization and recombination cooling}

Photoionization heating is accounted for via tracking the photoionization heating for each gas cell during the RT step. This means that the code explicitly accounts for differences in excess heating between soft and hard ionizing radiation, as long as the radiation field is well-sampled. Radiative and dielectronic recombination, collisional ionization, and charge exchange heating/cooling are included using the rates as described in Section~\ref{sec:ionization and recombination}. 

\subsubsection{Line transfer}
\label{sec:Line transfer}

In addition to recombination, collisional excitation of electrons can also induce line emission. To capture this process for H and He, we use analytic expressions, with \HI rates from \citet{Smith:2022aa} and \HeI and \HeII rates from \citet{Cen:1992aa}. Metal line emission is treated with a newly implemented level population solver. The atomic data is derived from the CHIANTI 11.0.2 database \citep{Dere:1997aa,Dere:2019aa,Dere:2023aa}. Our default network of metal ions for line emission includes \CI--\CVI, \NI--\NVII, \OI--\OVIII, \ion{Ne}{II}--\ion{Ne}{VIII}, \SII-\SIV, and \FeII--\ion{Fe}{VI}. Our default network uses the number of levels in the NIRVARNA code \citep{Ziegler:2005aa,Ziegler:2018aa}. For S and Ne, which are not included in the NIRVARNA network, we use 15 levels for each ion, which is the same number of levels used by Cloudy for non-Fe ions. The level population solver includes collisional excitation, collisional de-excitation, and spontaneous radiative decay. Stimulated excitation/emission due to the continuum could in principle be incorporated into the solver (by saving and reloading the relevant portion of the radiation field), but this feature is currently not used in \colt applications as it requires further development and testing. This is because full line emission RT is typically calculated separately, and so we would only follow stimulated excitation for a single transition at a time, which is not self-consistent and may have unintended consequences. We plan to implement fully self-consistent stimulated emission in the future, including both continuum and line pumping. A preliminary implementation of multi-level line MCRT in \colt is already being used for idealised applications.

\subsubsection{Free-free emission}
\label{sec:Free-free emission}

Free--free emission is implemented for \HII, \HeII, and \HeIII, each with an electron partner. The cooling due to free--free emission for an ion, $X$, is given by
\begin{equation}
\Lambda_\mathrm{ff}(T,n_X,n_e) = n_X \, n_e \frac{32\,Z^2 e^6}{h c^3} \left(\frac{2\pi^3 h T}{3^3 k_B m_e^3}\right)^{1/2}  \langle g_\mathrm{ff}(Z,T,\nu)\rangle \,,
\end{equation}
where $Z$ is the ion's charge, $e$ is the electron charge, $m_e$ is the electron mass, $h$ is the Planck constant, $c$ is the speed of light, and $k_B$ is the Boltzmann constant. $\langle g_\mathrm{ff}(Z,T,\nu)\rangle$ is the total, non-relativistic free--free Gaunt factor, which we take from \citet{van-Hoof:2014aa}.

\subsubsection{Compton scattering}
\label{sec:Compton scattering}

Compton heating and cooling are implemented for the cosmic microwave background (CMB) following the analytic expression from \citet{Haiman:1996aa}
\begin{equation}
  \Lambda_\mathrm{Compton}(T,a,n_e)=1.017\times 10^{-37} \,n_e\,\left( \frac{T_\mathrm{CMB}}{a}\right)^4  \left( T- \frac{T_\mathrm{CMB}}{a}\right) \,,
\end{equation}
where $a$ is the scale factor and $T_\mathrm{CMB}$ is the temperature of the CMB at $z=0$, which we assume to be 2.727\,K.

\subsection{Equilibrium solver}
\label{sec:Equilibrium solver}

After each RT step, the joint ionization/thermal equilibrium solver is called. If temperature solving is not desired, then only the ionization solver is used. In this case, either the simulation temperatures or internal energies are used. The temperature case is most simple; ionization states are updated iteratively to converge on states that balance the radiative, collisional, and charge exchange ionization and recombination rates. When internal energy is used, the procedure is similar except that the temperature is updated iteratively based on the internal energy and mean molecular weight, which is affected by the ionization state of the gas. Of course, the new ionization state of the gas impacts the transport of ionizing radiation. For this reason, we iteratively run RT steps until the ionization states reach convergence. For large galaxy simulations, we typically will reach an initial preconditioning convergence with $10^7$ photons before running again to convergence with $10^8$ or more photons.

The thermal equilibrium solver adds an additional layer of complexity. After the RT step has completed, the ionization solver is called to find the equilibrium ionization states at the initial temperature. We then find the cooling rate at that given temperature, accounting for all the previously described processes. If the cooling rate for a given cell is not close to zero and it is not tending to heat (cool) below (above) a temperature limit, then we proceed to find the equilibrium temperature. This is essentially a root finding exercise where we are aiming to balance the heating and cooling rates. Due to the fact that the cooling rates can be quite complex and are affected by the changing ionization states of the gas with temperature, we use the simple and robust bisection method to converge on the equilibrium temperature. At each new temperature, the ionization states and cooling rates are calculated, and the temperature range above or below that value is eliminated. This process repeats iteratively until the temperature changes by a value lower than the user-specified tolerance.

The new temperature is not directly set to the equilibrium temperature. Instead, we use a gain factor between 0.1 and 1, inspired by that used in the \textsc{sirocco} code \citep{Matthews:2025aa}. The gain factor decreases when the temperature is oscillating around the solution, and increases otherwise. This helps to avoid large oscillations around the true temperature and aids convergence. For large galaxy simulations, we typically will first run to an initial convergence with the ionization solver only using $10^7$ photons. We then run with the thermal equilibrium solver to convergence with $10^7$ photons, and again to more stringent convergence with $10^8$ photons. We have found that this approach achieves convergence with the minimal number of RT steps, which is the computationally limiting factor. The number of photons and strategy used is user-defined and should vary with the size of the simulation and the desired degree of convergence. Convergence is user-defined based on the fractional change in the global rate of hydrogen recombination, although in the future we plan to include additional targeted criteria and implement an adaptive convergence scheme similar to that of \textsc{arepo-mcrt} \citep{Smith:2020aa}.

In galaxy simulations, it is not necessarily advantageous to bring all gas into thermal equilibrium. This is because our cooling scheme cannot account for hydrodynamic and non-equilibrium heating and cooling. To address this issue, cooling can be limited by explicitly implementing a maximum cooling time (e.g., 1\,Myr), or it can be specified based on the sound-crossing time, $t_\mathrm{sound}$. Our default setting is the Courant sound crossing time, which is the sound crossing time multiplied by a Courant factor, $C_\mathrm{CFL}$, of 0.2 as a conservative default choice. The maximum cooling in a cell is then given by
\begin{equation}
\Delta T_\mathrm{max}=\frac{\text{d}T}{\text{d}t}\, C_\mathrm{CFL}\, t_\mathrm{sound} \, ,
\end{equation}
where $\text{d}T/\text{d}t$ is the cooling rate. This means that gas can cool to a temperature if it could do so before hydrodynamic forces can react. In practice, this means that cooler, denser gas can cool more, whereas hotter and more diffuse gas is less able to cool. This preserves diffuse gas, particularly in the CGM and IGM, while allowing the code to fully treat photoionized gas in the ISM. We do not limit the temperature to which a cell can be heated because the radiation field is more accurate (both spatially and spectrally) using our MCRT solver compared to the on-the-fly radiation field.

\begin{figure*}
    \centering
	\includegraphics[width=\textwidth]{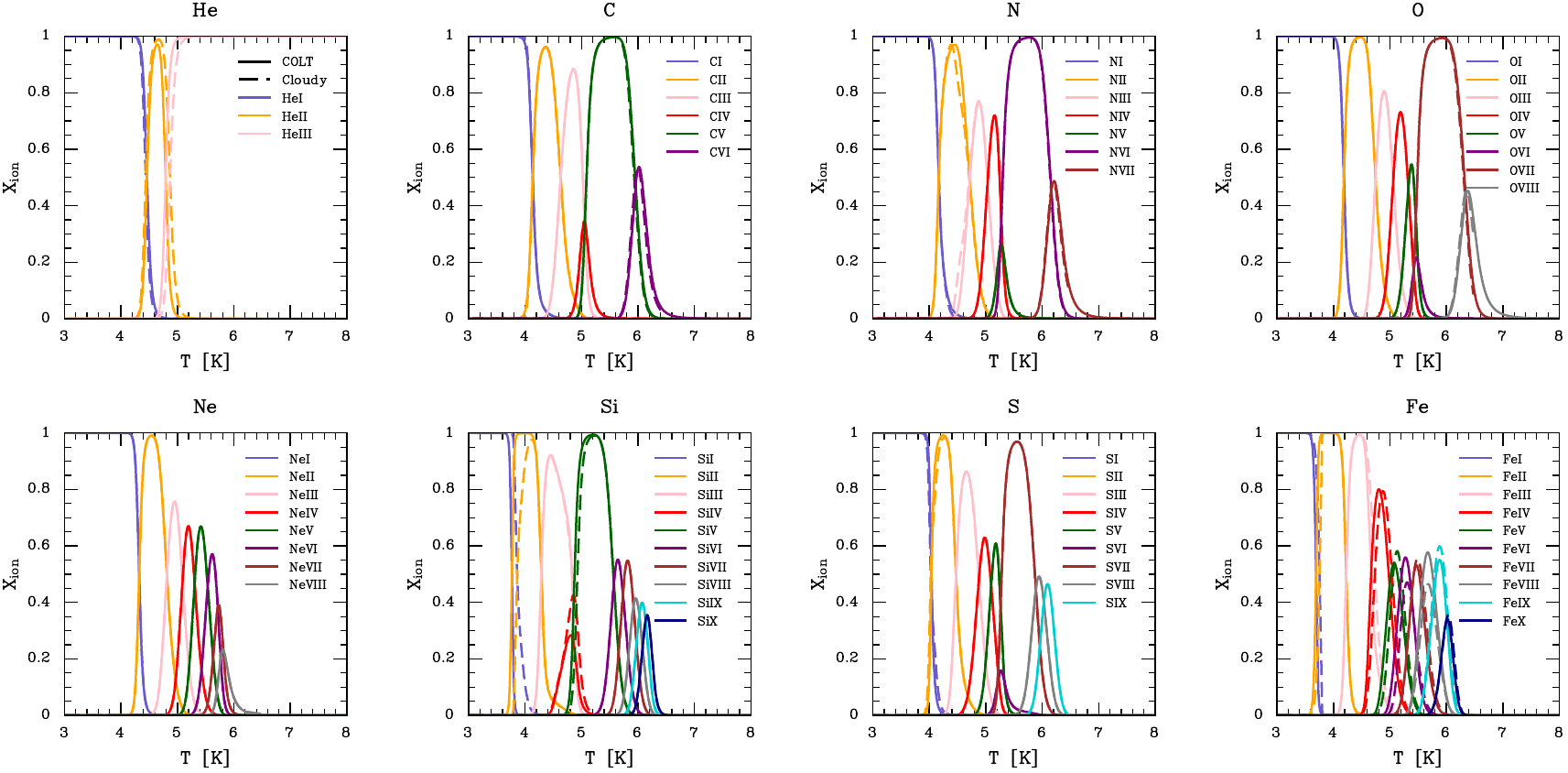}
    \caption{The ionization states of gas in collisional ionization equilibrium (CIE) as a function of temperature for distributions He, C, N, O, Ne, Si, S, and Fe from \colt (solid) and \textsc{Cloudy} (dashed), evaluated from $T=10^3$\,K to $T=10^8$\,K at a fixed density of $\mathrm{n_H}=100\,\mathrm{cm^{-3}}$. Photoionization, molecules, dust, and induced processes are disabled. The agreement between the codes is overall excellent across the range of temperatures. Fe shows the largest overall offset, with a mean absolute error of 6.1\%, and the mean absolute error for other ions is $\lesssim$5\% (see text).}
    \label{fig:CIE_test}
\end{figure*}

\subsection{Observable emission}
\label{sec:Observable emission}

Non-ionizing, observable emission is included in \colt for stellar continuum, nebular continuum, and nebular line emission. The RT carried out for non-ionizing emission is largely the same as described in Section~\ref{sec:Radiative transfer}, although photoionization cross-sections are not included below the respective ionization thresholds. There are additional peculiarities for the RT of resonant lines such as the Ly$\alpha$ line of neutral hydrogen, which \colt was originally designed to target and are described in \citet{Smith:2015aa, Smith:2019aa, Smith:2022aa, Smith:2025aa}.

\subsubsection{Ionizing radiation sources}
\label{sec:Ionizing radiation sources}

Ionizing radiation can be emitted from several different types of sources. The dominant source in simulations of galaxies is typically stars, and we favour stellar SEDs derived from the Binary Population and Spectral Synthesis (BPASS) model \citep[v2.2.1;][]{Eldridge:2009aa, Eldridge:2017aa}\footnote{Further information on BPASS can be found on the project website at \href{https://bpass.auckland.ac.nz}{\texttt{bpass.auckland.ac.nz}}.}, however we have also used \colt with the SEDs presented in \citet{Bruzual:2003aa}. \colt can be supplied with input SED files to make use of any desired stellar library. We note that other sources of ionizing radiation are included in \colt, such as blackbody, AGN, UVB, or tabulated spectra.

\subsubsection{Nebular continuum}
\label{sec:Nebular continuum}

We have newly implemented nebular continuum emission into \colt, including free--free, free--bound, and two-photon emission. We largely follow the rates and approach used in the \textsc{nebular} code \citep{Schirmer:2016aa}. We use rates from \citet{Ercolano:2006aa} to calculate the frequency and temperature-dependent free--bound emissivity for \HI, \HeI, and \HeII. 

Free--free emission is included between electron partners and \HII, \HeII, and \HeIII. The free--free emissivity at frequency $\nu$ is given by
\begin{equation}
j_\mathrm{ff}(\nu) = n_X \, n_e \frac{8\,Z^2 e^4 h}{3 m_e^2 c^3} \left(\frac{h \nu_0}{3 \pi k_B T}\right)^{1/2} \exp\left(\frac{-h \nu}{k_\text{B} T}\right) \langle g_\mathrm{ff}(Z,T,\nu)\rangle \,,
\end{equation}
where $h\nu_0$ is the ionization energy for \HI. $\langle g_\mathrm{ff}(Z,T,\nu)\rangle$ is the non-relativistic, thermally-averaged Gaunt factor, which we take from \citet{van-Hoof:2014aa}. Note that this is \textit{not} equivalent to the total thermally-averaged Gaunt factor used in Section~\ref{sec:Free-free emission}, which is averaged over frequency.

Two-photon emission arises from the transition of an electron from the $2^2\mathrm{S}$ to the $1^2\mathrm{S}$ state. This is a forbidden transition, so rather than emitting a single photon with a discrete energy, two photons are emitted which share energy probabilistically, generating a characteristic continuum emission which peaks at one-half of the ion's Ly$\alpha$ frequency, $\nu_{12}/2$. The emissivity of the two-photon continuum is strongly dependent on density due to angular momentum changing ($l$-changing) collisions changing the electron from the $2^2\mathrm{S}$ to the $2^2\mathrm{L}$ state, from which it can promptly decay via Ly$\alpha$ emission. It is for this reason that the two-photon continuum is heavily suppressed at $\mathrm{n_H}\gtrsim1000\,\mathrm{cm^{-3}}$ \citep{Schirmer:2016aa}. We include two-photon emission from \HI and \HeII. We do not include \HeI two-photon emission, which is a departure from the \textsc{nebular} implementation, due to the poor availability of data and because it is expected to be orders of magnitude less intense than \HI two-photon emission \citep{Schirmer:2016aa}. The frequency-dependent two-photon emissivity is given by
\begin{equation}
\begin{aligned}
j_{2q}(\nu,n_e,n_X,T) &= \\
n_X \, n_e \, &\alpha^\mathrm{eff}_{2^2\mathrm{S}}(T,n_e)\,
\frac{h\nu\,A(\nu/\nu_{12})}{4\pi \,A_{2q}\,\nu_{12}}
\left(1+\frac{\sum_Y n_Y\,q^Y}{A_{2q}}\right)^{-1} \,,
\end{aligned}
\end{equation}
where $\alpha^\mathrm{eff}_{2^2\mathrm{S}}$ is the effective recombination rate to the $2^2\mathrm{S}$ state, which we take from \citet{Hummer:1987aa}. $A_{2q}$ is the total transition probability and $A(\nu/\nu_{12})$ is the frequency-dependent transition probability, which are both calculated following \citet{Nussbaumer:1984ab}. The sum $\sum_Y n_Y\,q^Y$ represents the rate of $l$-changing collisions due to different collisional partners, $Y$. We include $l$-changing collisions due to both $n_e$ and \HII collisional partners, calculating the collision rate transition coefficient for each following the \citet{Pengelly:1964aa} formalism.

\subsubsection{Nebular line emission}
\label{sec:Nebular line emission}

H and He lines are calculated assuming Case B conditions with line-specific analytic formulae. We use the total recombination rates from \citet{Hui:1997aa}, which are the same as those employed by the ionization solver. We then use the probability of emitting a given line (e.g. H$\alpha$) per recombination even to compute the emissivity of the line. For a given H or He line, denoted as $X$, the luminosity in a given volume is given by
\begin{equation}
  L_X^\mathrm{} = h \nu_X \int \left(P_{X}(T,n_{\rm e}) \, \alpha(T) \, n_\text{ion+1} + q_{\mathrm{col},\,X} \, n_\text{ion}\right)\,n_{\rm e} \,\text{d}V \, ,
\end{equation}
where $h\nu_X$ is the line energy, $\alpha$ is the total recombination coefficient, and $P_{X}$ is the probability for that emission line photon to be emitted per recombination event. The collisional contribution to the line is controlled by $q_{\mathrm{col},\,X}$, which is the collisional excitation rate coefficient. We calculate $q_{\mathrm{col},\,X}$ for \HI lines using the formulae presented in \citet{Smith:2022aa}.
Previously, we have calculated metal line emission using simple functions \citep{McClymont:2024aa}, which is less accurate and necessitates a more limited library of lines compared to our newly implemented level population solver. We use atomic data from the CHIANTI 11.0.2 database \citep{Dere:1997aa,Dere:2019aa,Dere:2023aa}.
Resonance lines are sourced as described above, but require special treatment for their partial frequency redistribution. A full discussion of the implementation of line RT can be found in \citet{Smith:2015aa}, with various updates to extend the applicability or efficiency.

\subsection{Outputs}
\label{sec:Outputs}

\colt produces an assortment of useful outputs. This includes direct photon outputs as probes of the MCRT, star particle and cell-level data to understand the sources and sinks, such as the ionization states, the heating/cooling rates, emissivity, and radiation pressure and forces, as well as observational data probes, such as escape and absorption fractions for globally-integrated, camera-based next-event estimation, and per solid angle or per-galaxy histograms. This data is well-suited to studying the physical processes. \colt is also designed to produce outputs which can be compared to observations, including emission maps, integral-field spectroscopy (IFS), integrated emission, radially binned emission, and half-light sizes. The specific outputs and their properties (e.g., resolution) are user-defined and depend on the simulation being analysed and the goals of the analysis. For example, it may not be appropriate to store the same resolution IFS outputs for a large cosmological simulation with many subhalos as for the analysis of an isolated galaxy simulation.

\begin{figure}
    \centering
	\includegraphics[width=\columnwidth]{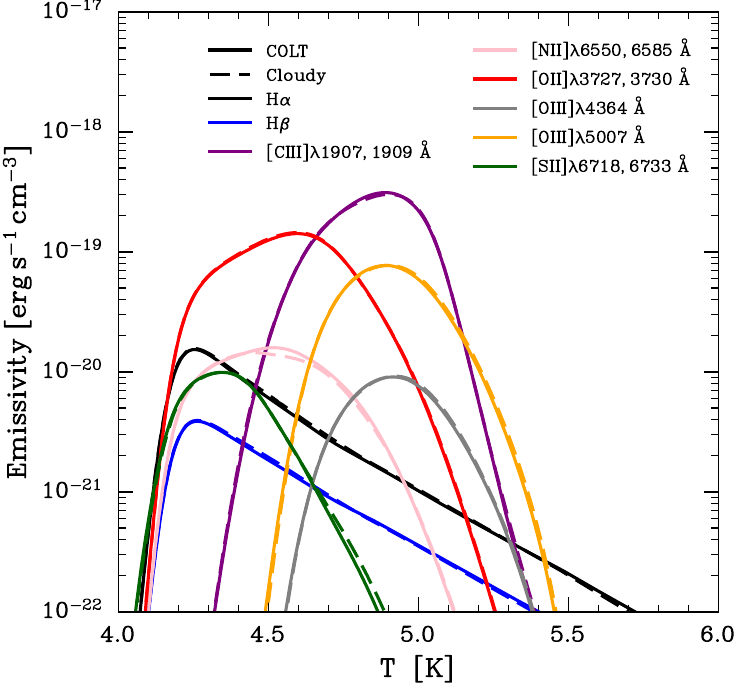}
    \caption{Emissivity versus temperature for a set of nebular emission lines, computed in the same CIE set-up as Fig.~\ref{fig:CIE_test} with \colt (solid) and \textsc{Cloudy} (dashed). We show \ion{C}{III}]$\lambda$1907,1909~\AA, [\ion{N}{II}]$\lambda$6550,6585~\AA, [\ion{O}{II}]$\lambda$3727,3730~\AA, [\ion{O}{III}]$\lambda$4364~\AA, [\ion{O}{III}]$\lambda$5007~\AA, and [\ion{S}{II}]$\lambda$6718,6733~\AA, all of which have emissvities calculated with our newly implemented level population solver. Agreement is excellent across the full temperature range, demonstrating the accuracy of our level population solver, with mean absolute fractional errors of $\lesssim$3.5\% for all lines (see text). We also show the emissivities of H$\alpha$ and H$\beta$, which are calculated using tables assuming the Case B approximation \citep{Storey:1995aa}, and show similarly good agreement (see text).}
    \label{fig:CIE_lines}
\end{figure}

\section{Model validation}
\label{sec:Model validation}

\subsection{Collisional ionization equilibrium}
\label{sec:Collisional ionization equilibrium}

To benchmark the ionization solver in the absence of radiation, we compute collisional ionization equilibrium (CIE) ionization fractions for 8 elements (He, C, N, O, Ne, Si, S, Fe) as a function of temperature. We initialized a grid of cells with steadily increasing temperature from $10^3$\,K to $10^8$\,K. We disable photoionization, molecules, dust, and induced processes, and evolve only collisional ionization, radiative and dielectronic recombination, and charge exchange. We set the Cloudy collisional rates to the same as used in \colt \citep{Voronov:1997aa}. We apply a fixed density of $\mathrm{n_H}=100\,\mathrm{cm^{-3}}$ and use the default Cloudy \ion{H}{II} region abundances \citep{Baldwin:1991aa,Rubin:1991aa,Rubin:1993aa,Osterbrock:1992aa}. For our reference solution, we run a Cloudy model with a similar setup to ours. We show the results for both \colt and Cloudy in Fig.~\ref{fig:CIE_test}. The two codes show excellent agreement across the full temperature range for each element, demonstrating the accuracy of the ionization solver. We calculated the mean absolute error by averaging the fractional difference between the most abundant ion calculated by Cloudy and the abundance of that ion calculated with \colt. We average in logspace temperature across all temperatures where there is at least a 5\% abundance of a tracked ion. The mean absolute errors are as follows: He (3.0\%), C (0.1\%), N (1.1\%), O (0.4\%), Ne (0.1\%), Si (5.1\%), S (0.01\%), and Fe (6.1\%).

In order to test our level population solver and resulting line emission, we compute the emissivity of a selection of metal lines, which are selected as examples of important coolants and/or observational indicators. The selected emission lines are as follows: \ion{C}{III}]$\lambda$1907,1909~\AA, [\ion{N}{II}]$\lambda$6550,6585~\AA, [\ion{O}{II}]$\lambda$3727,3730~\AA, [\ion{O}{III}]$\lambda$4364~\AA, [\ion{O}{III}]$\lambda$5007~\AA, and [\ion{S}{II}]$\lambda$6718,6733~\AA. The luminosity densities are calculated with both \colt and Cloudy in the CIE setup described above. We present the results in Fig.~\ref{fig:CIE_lines}, showing excellent agreement between the codes. We also show the emissivities of H$\alpha$ and H$\beta$, which are calculated with tables assuming Case B recombination \citep{Storey:1995aa}, and these lines also agree well with Cloudy. We calculate the mean absolute fractional error, $\langle|L_\mathrm{Cloudy} - L_\mathrm{COLT}| / (L_\mathrm{Cloudy} + L_\mathrm{COLT})\rangle$, across all temperatures where the emissivity of the line is at least $10^{-22}$\,erg\,s$^{-1}$\,cm$^{-3}$. The mean absolute fractional errors are as follows: \ion{C}{III}]$\lambda$1907,1909~\AA\ (1.9\%), [\ion{N}{II}]$\lambda$6550,6585~\AA\ (2.3\%), [\ion{O}{II}]$\lambda$3727,3730~\AA\ (1.0\%), [\ion{O}{III}]$\lambda$4364~\AA\ (1.8\%), [\ion{O}{III}]$\lambda$5007~\AA\ (3.5\%), and [\ion{S}{II}]$\lambda$6718,6733~\AA\ (3.3\%).

\subsection{Str\"omgren sphere}
\label{sec:Stromgren sphere}

\begin{figure}
    \centering
	\includegraphics[width=\columnwidth]{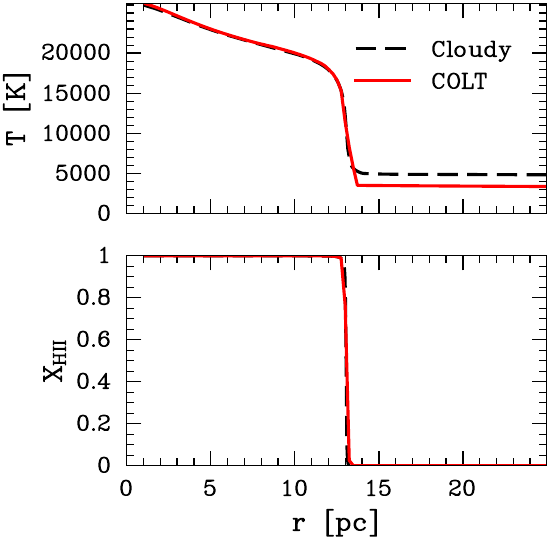}
    \caption{Spherically symmetric H-only Str\"omgren test comparing \colt (solid) and \textsc{Cloudy} (dashed). The ionizing source is a BPASS v2.2.1 SED for an IMF-averaged 5\,Myr stellar population with a metallicity of $Z=0.02$ and mass of $10^{4.7}\,\mathrm{M_\odot}$. We show the equilibrium temperature (\textit{upper panel}) and the fractional abundance of hydrogen ions as a function of radius (\textit{lower panel}). The agreement between the codes is excellent, with both codes recovering a sharp ionization front and decreasing temperature with radius. Outside the ionized zone, temperatures diverge because \colt’s network currently lacks molecular cooling, which suppresses further cooling. This difference is not relevant for the minimally metal-enriched ISM regime which we target.}
    \label{fig:stromgren_hydrogen_test}
\end{figure}

\begin{table}
    \centering
    \begin{tabular}{ccccc}
        \hline
        \multicolumn{5}{|c|}{Str\"omgren sphere emission lines} \\
        \hline
        Ion & Wavelength & $L_\mathrm{COLT}$ & $L_\mathrm{Cloudy}$ & $|\Delta$|\\
         &  [\AA] & [$10^{37}$ erg\,s$^{-1}$] & [$10^{37}$ erg\,s$^{-1}$] &  \\
        \hline
        \HI & 4863 & $100$ & $103$ & $0.01$ \\
        \HI & 6565 & $294$ & $290$ & $0.01$ \\
        \CIII & 1907,1909 & $45.3$ & $39.6$ & $0.07$ \\
        \NII & 6550,6585 & $21.3$& $18.2$ & $0.08$ \\
        \OII & 3727,3730 & $62.0$ & $56.2$ & $0.05$ \\
        \OIII & 4364 & $3.38$ &$3.06$ & $0.05$ \\
        \OIII & 5007 & $572$ & $541$ & $0.03$ \\
        \SII & 6718,6733 & $13.9$ & $11.5$ & $0.09$ \\
        \hline
    \end{tabular}
    \caption{Integrated nebular line luminosities from the Str\"omgren sphere benchmark, comparing \colt against \textsc{Cloudy} for the same geometry. The setup is the same as for Fig.~\ref{fig:stromgren_test}. Columns list the ion from which a line arises, the vacuum wavelength, the total line luminosity predicted by \colt and by \textsc{Cloudy}, and the absolute fractional difference $|\Delta|=|L_\mathrm{Cloudy} - L_\mathrm{COLT}| / (L_\mathrm{Cloudy} + L_\mathrm{COLT})$. Overall, the agreement between the codes is good, with errors at the $\lesssim10\%$ level.}
    \label{tab:line_ratios}
\end{table}

\begin{figure*}
    \centering
	\includegraphics[width=\textwidth]{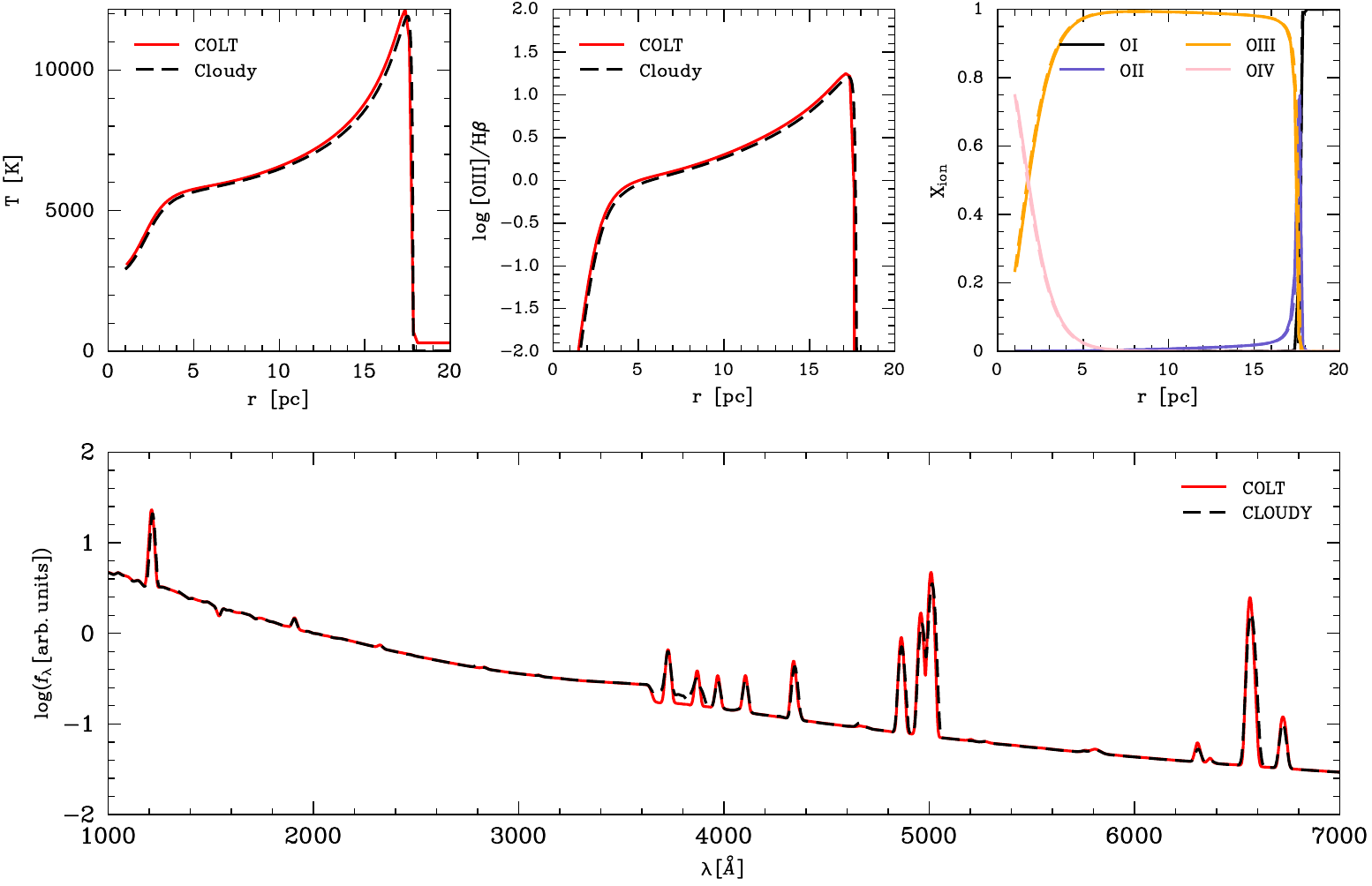}
    \caption{Metal-enriched Str\"omgren test comparing \colt (solid) and \textsc{Cloudy} (dashed). We use the same setup as for Fig.~\ref{fig:stromgren_hydrogen_test}, except we use \HII region abundances (apart from He) and we change the ionizing source to a BPASS v2.2.1 SED for an IMF-averaged 3\,Myr stellar population with a metallicity of $Z=0.006$. We show the equilibrium temperature (\textit{upper left panel}), fractional abundance of oxygen ions (\textit{upper central panel}), and [\ion{O}{III}]$\lambda$5007~\AA\,/ H$\beta$ ratio (\textit{upper right panel}) as a function of radius. In the lower panel, we show the intrinsic spectrum of the Str\"omgren sphere, which includes high-EW emission lines and a Balmer jump. The agreement between the codes is excellent, which shows the combined accuracy of our ionization solver, radiation transport, thermal equilibrium solver, level population solver, nebular continuum, and cooling/heating rates.
    }
    \label{fig:stromgren_test}
\end{figure*}

In order to test our coupled ionization and thermal equilibrium solvers, we simulate Str\"omgren spheres. We conduct this test using spherical geometry, within an empty inner radius of 1\,pc and a maximum radius of 25\,pc. We disable molecules, dust, and induced processes. In Cloudy, we also enforce Case B (including the local absorption of diffuse ionizing fields arising from the gas) and switch off induced processes and molecules. The \colt cells have a length of 0.25\,pc ($\sim$52 cells in the ionized region), and Cloudy cells are determined by the code's adaptive algorithm.

We first conduct a test of pure H gas. For the central source, we use a stellar SED derived from BPASS v2.2.1 \citep{Eldridge:2009aa, Eldridge:2017aa}, representing a $10^{4.7}\,\mathrm{M_\odot}$ IMF-averaged stellar population with a metallicity of $Z=Z_\odot=0.02$ and age of 5\,Myr. We show the results of this test in Fig.~\ref{fig:stromgren_hydrogen_test}, where we can see that both the temperature and hydrogen ionization fraction agree well between the codes. Outside of the ionized region, the temperatures between the codes diverge due to the fact that we do not currently have a molecular network which allows pure hydrogen gas to cool effectively below a few thousand K. This is not a concern for the regime in which we will plan to operate the code, as we are mainly concerned with nebular emission arising from the ISM, which is always at least minimally metal-enriched and which has a non-negligible UVB. However, this would be important for modelling the transport of Ly$\alpha$ through the IGM in the early Universe, where the gas may be pristine and have an extremely low ionization background.  Future development of the code will either expand the network to include molecular cooling, or alternatively flag cool, pristine IGM gas to avoid treating it with the thermal equilibrium solver. However, we note that the ISM escape of Ly$\alpha$ can still be accurately treated with the current model, given that the ISM will be at least minimally enriched with metals and the ionizing radiation field will be more significant.

\begin{figure*}
    \centering
	\includegraphics[width=\textwidth]{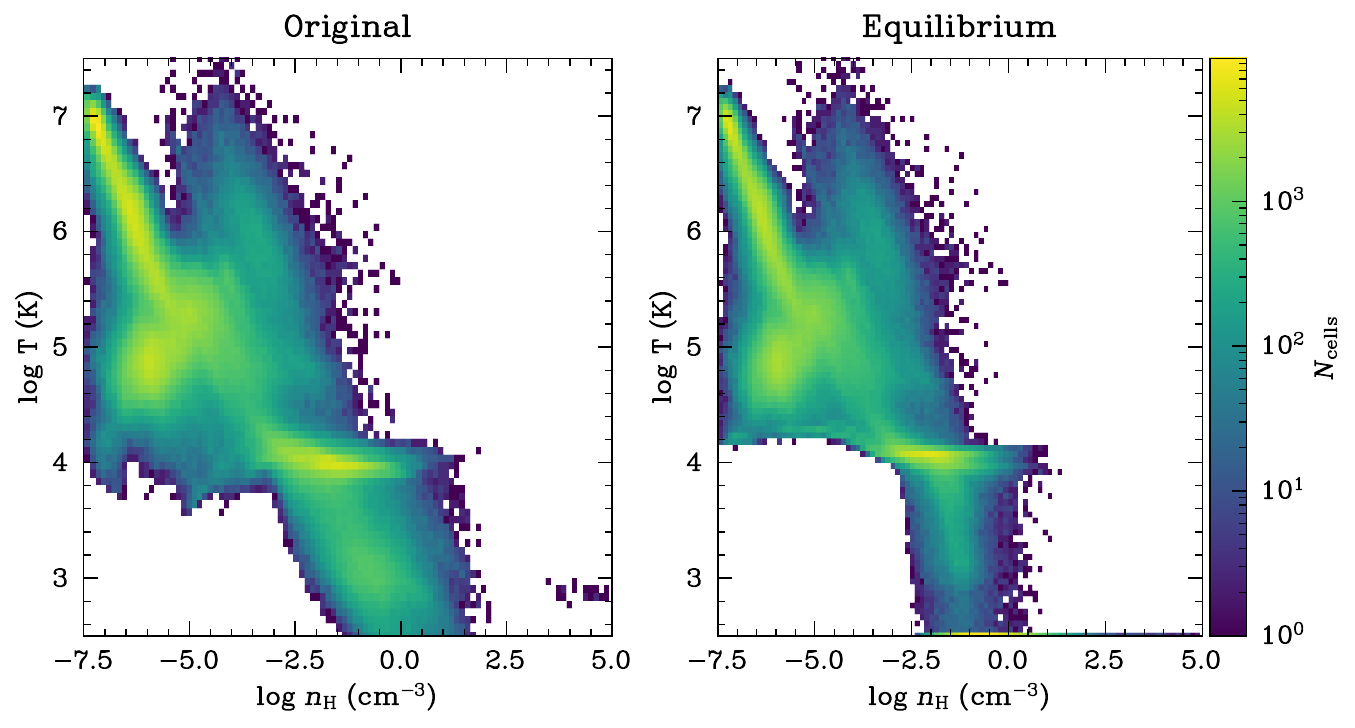}
    \caption{Temperature-density phase space of the LMC-like simulation before and after post-processing with \colt. Left: the original on-the-fly temperatures recorded by the hydrodynamic run. Right: the temperatures obtained after running \colt with the joint ionization and thermal equilibrium solver, including Courant-limited cooling and the same UVB as in the simulation. The distribution of gas cells shifts as expected, with hotter ($T\gtrsim10^4$\,K), lower-density ($n_\mathrm{H}\lesssim1\,\mathrm{cm^{-3}}$) gas outside the ISM being largely preserved, while cooler, denser ISM gas shifts toward equilibrium. An immediate consequence is the reduction of gas occupying the intermediate temperatures ($T=10^3-10^4$\,K), with a clearer separation between warm ionized ($T=10^4$\,K) and cold neutral ($T\lesssim10^3$\,K) phases. Despite also existing outside the IGM, extremely diffuse gas ($n_\mathrm{H}\lesssim0.01\,\mathrm{cm^{-3}}$) initially near $T\lesssim10^4$\,K is modestly heated by the UVB because we do not limit heating. Together, these changes demonstrate that the equilibrium solver corrects temperature systematics relevant for nebular-line emissivities, such as an excessive amount of lukewarm gas, while leaving the CGM structure intact.}
    \label{fig:temp_dens_comparison}
\end{figure*}

We also conduct another Str\"omgren sphere test with the default \HII region abundances in Cloudy. He was not included in this test because Cloudy Case B options do not apply fully to He, and can lead to excessively large He ionization from excited states. For the central source, we again use a stellar SED derived from BPASS representing a $10^{4.7}\,\mathrm{M_\odot}$ IMF-averaged stellar particle, although we use a lower metallicity of $Z=0.006$ and an age of 3\,Myr. We retain the rest of the physical setup from the original test, including 0.25\,pc cell length ($\sim$72 cells in the ionized region). In Fig.~\ref{fig:stromgren_test} we show the results for this Str\"omgren sphere model. The upper left panel shows the temperature as a function of radius as calculated by each code. There is some slight discrepancy between the codes, with Cloudy finding a volume-averaged temperature of 8223\,K in the ionized region compared to 8484\,K for \colt. We note that this discrepancy is because in calculating the temperature with \colt, we switched off the cooling contribution for \ion{Ne}{III} and \ion{S}{III}. This is because the emission from these ions is impacted by line transfer effects. If we turn on these ions in \colt and switch off line transfer in Cloudy, we find near exact agreement in temperature; however, this also impacts the Balmer line emission from Cloudy (this appears to be an inconsistency in Cloudy as we are enforcing Case B). In any case, the agreement with Cloudy is overall excellent for the physical processes that are included in the code.

The fact that we successfully converged on the correct temperatures implies that our level population solver, and therefore metal line emission, is largely accurate. This was also shown in the CIE case in Fig.~\ref{fig:CIE_lines}. However, to further demonstrate the accuracy, in Fig.~\ref{fig:stromgren_hydrogen_test} we show the O ionization fractions and the [\ion{O}{III}]/H$\beta$ line ratio throughout each region of the gas. This agreement is excellent, with the ratio largely tracking temperature throughout the cloud. We show the results for the total line emission from both \colt and Cloudy for a variety of emission lines in Tab.~\ref{tab:line_ratios}. Overall, agreement is generally good between the two codes, with errors at the $\lesssim10\%$ level. Lines with larger errors tend to be highly temperature sensitive (e.g. [\ion{O}{III}]$\lambda$4364~\AA) or arise from ions with high ionization potentials (e.g. \ion{C}{III}]$\lambda$1908~\AA). The errors are primarily driven by the small discrepancies in the temperature of the gas discussed previously. 

In the lower panel of Fig.~\ref{fig:stromgren_hydrogen_test}, we show the intrinsic spectrum of the \HII region produced by \colt and Cloudy, which both feature a prominent Balmer jump. The mean absolute fractional error, $\langle|f_{\lambda,\mathrm{Cloudy}} - f_{\lambda,\mathrm{COLT}}| / (f_{\lambda,mathrm{Cloudy}} + f_{\lambda,mathrm{COLT}})\rangle$, is 4\%. Other than the previously discussed small errors in the line strengths, another source of error is the continuum redwards of the Balmer jump. A clear offset between the codes can be seen at $\sim$3800\AA, which is due to the fact that we do not currently include the many higher-order Balmer lines which blend into each other, causing excess emission in this region.

\section{Large Magellanic Cloud-like simulation}
\label{sec:LMC Simulation}

\begin{figure}
    \centering
	\includegraphics[width=\columnwidth]{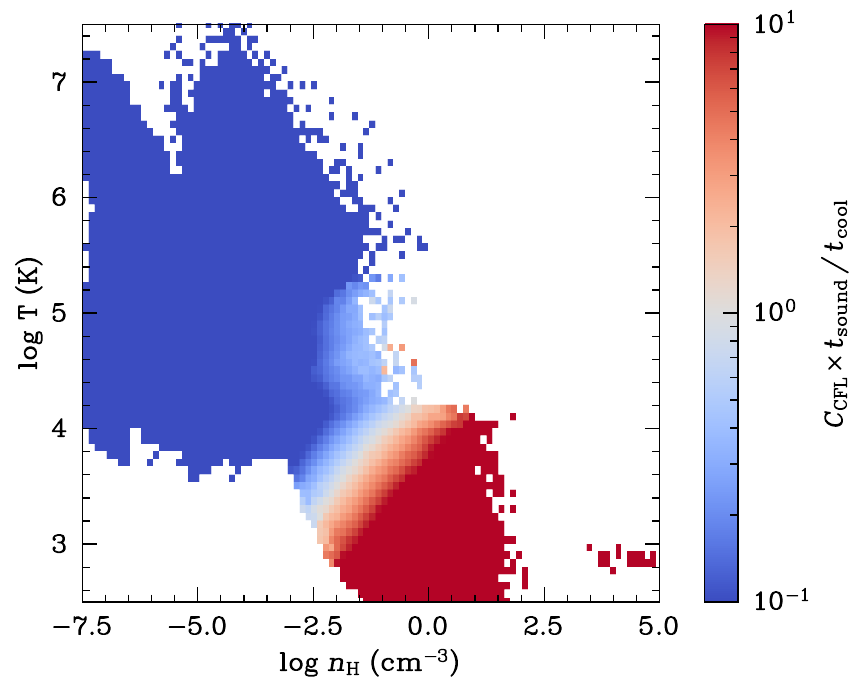}
    \caption{Temperature--density phase space in the LMC-like simulation colored by the sound-crossing time divided by the cooling time, $C_\mathrm{CFL}\times t_\mathrm{sound}/t_\mathrm{cool}$, where $C_\mathrm{CFL}$ is the Courant factor, for which we adopt a value of 0.2. Blue indicates $C_\mathrm{CFL}\times t_\mathrm{sound}\gg t_\mathrm{cool}$ and red indicates $C_\mathrm{CFL}\times t_\mathrm{sound}\ll t_\mathrm{cool}$. The net cooling rate calculated with \colt is used to calculate $t_\mathrm{cool}$, and $t_\mathrm{sound}$ is calculated based on each cell's sound speed and its radius. This diagnostic explains the behaviour seen in Fig.~\ref{fig:temp_dens_comparison} and motivates the Courant-limited cooling strategy adopted here, where the maximum negative change in temperature calculated in post-processing with \colt is limited to $\frac{\text{d}T}{\text{d}t}\times C_\mathrm{CFL}\times t_\mathrm{sound}$.}
    \label{fig:courant_limiting}
\end{figure}

\subsection{Simulation details}
\label{sec:LMC Simulation details}

In order to test our code on galactic scales, we use a simulation of an isolated Large Magellanic Cloud-like galaxy ($M_\mathrm{halo} = 1.1 \times 10^{11}\,\mathrm{M_\odot}$), which was presented in \citet{Kannan:2020aa,Kannan:2021aa}. The simulations employ \textsc{AREPO-RT} \citep{Kannan:2019aa}, the radiation-hydrodynamics extension of the moving-mesh code \textsc{AREPO} \citep{Springel:2010aa,Weinberger:2020aa}. The initial conditions for the simulation consist of a dark matter halo, a stellar bulge, a stellar disk, and a gaseous disk, and was set up as described in \citet{Hernquist:1993aa,Springel:2005aa}. The stellar and gas disks follow exponential profiles with scale lengths of 2.8\,kpc and 1.4\,kpc, respectively. The vertical profile of the stellar disk is sech$^2$ with a scale height of 140\,pc. Stars are initialized with an age of 5\,Gyr to avoid contribution to the ionization of the gas. The gas is distributed in hydrostatic equilibrium, with an initial gas fraction of 0.18 and an initial temperature of $10^4$\,K. The metallicity is set to $0.5\,\mathrm{Z_\odot}$ and new metal production is turned off to avoid excessive enrichment given the isolated nature of the simulation. The simulation box size is 200\,kpc and the simulation was run for 1\,Gyr. The gas cells have a mass resolution of $1.4\times10^{3}\,\mathrm{M_\odot}$ and the stars have a mass resolution of $2.8\times10^{3}\,\mathrm{M_\odot}$. The gravitational softening lengths are 3.6\,pc for gas and 7.1\,pc for stars.

The simulation employs the Stars and MUltiphase Gas in GaLaxiEs (SMUGGLE) star formation and feedback model \citep{Marinacci:2019aa,Kannan:2020aa}. Gas can cool down to a lower limit of $T_\mathrm{min}=10$\,K, with the cooling network being comprised of primordial H/He (atomic and molecular) processes, equilibrium metal-line cooling, gas-dust coupling, and photoelectric heating, and photoheating, all coupled self-consistently through the on-the-fly radiative-transfer solver \citep{Marinacci:2019aa,Kannan:2020aa}. Stars form stochastically from cold, self-gravitating gas above a threshold of $n_\mathrm{H}=1000\,\mathrm{cm^{-3}}$. Feedback from young and evolved stars is modelled via radiation coupled to the on-the-fly-solver (photoheating, radiation pressure, photoelectric heating), stellar winds from O/B and AGB stars, and supernovae. In addition to radiation sourced from stars, the simulation also includes a UVB \citep{Faucher-Giguere:2009aa} with a self-shielding prescription \citep{Rahmati:2013aa}. Dust physics follows the self-consistent formation and destruction framework of \citet{McKinnon:2016aa,McKinnon:2017aa}. Dust is produced by SN II, SN Ia, and AGB stars \citep{Dwek:1998aa}, grows by accretion of gas-phase metals \citep{Dwek:1998aa}, and is destroyed by supernova shocks and thermal sputtering in hot gas \citep{McKee:1989aa}.

\subsection{COLT}
\label{sec:LMC COLT}

We run \colt with our standard approach on a snapshot 1\,Gyr after the simulation has been initialized. Starting with the on-the-fly H and He ionization fractions, we run the code to initial convergence with the ionization solver only using $10^7$ photons. We then proceed to run \colt with the thermal equilibrium solver switched on, first running to convergence with $10^7$ photons, and then again with $10^8$ photons. During our runs with \colt, we use the Courant-limited cooling option to avoid over-cooling diffuse gas. We also include a UVB \citep{Faucher-Giguere:2009aa}, which is consistent with that used in the simulation. In Fig.~\ref{fig:temp_dens_comparison}, we show the original temperature--density phase space of the simulation compared to the newly recalculated phase space with \colt. 

The hotter ($T\gtrsim10^4$\,K) and lower-density ($n_\mathrm{H}\lesssim1\,\mathrm{cm^{-3}}$) gas which primarily lies outside the ISM remains largely intact. This is due to the Courant-limited cooling we have employed. In Fig.~\ref{fig:courant_limiting}, we show the ratio of the sound-crossing time to the cooling time. The hotter and lower-density gas has sound-crossing times are far shorter than the cooling time, leaving them largely unaffected by the thermal equilibrium solver. Because we do not limit the heating of gas, the coolest extremely diffuse gas which was initially at $T\approx10^4$\,K has been heated, primarily by the UVB. 

On the other hand, cooler ($T\lesssim10^4$\,K) and higher-density ($n_\mathrm{H}\gtrsim1\,\mathrm{cm^{-3}}$) gas shows sound-crossing times which are significantly longer than the cooling times, meaning that the temperature in this regime is determined entirely by the thermal equilibrium solver. A key feature of the new phase-space diagram is that there is significantly less gas between $10^3$--$10^4$\,K. Much of this gas is the spurious, partially ionized, temporarily unresolved Str\"omgren spheres, which causes unphysical line emission when the simulation temperature is used directly. While corrections can be applied for recombination lines \citep{Smith:2022aa}, such an approach remains problematic for highly temperature-dependent collisional lines. 

\subsection{Mock observations}
\label{sec:Mock observations}

\begin{figure*}
    \centering
	\includegraphics[width=\textwidth]{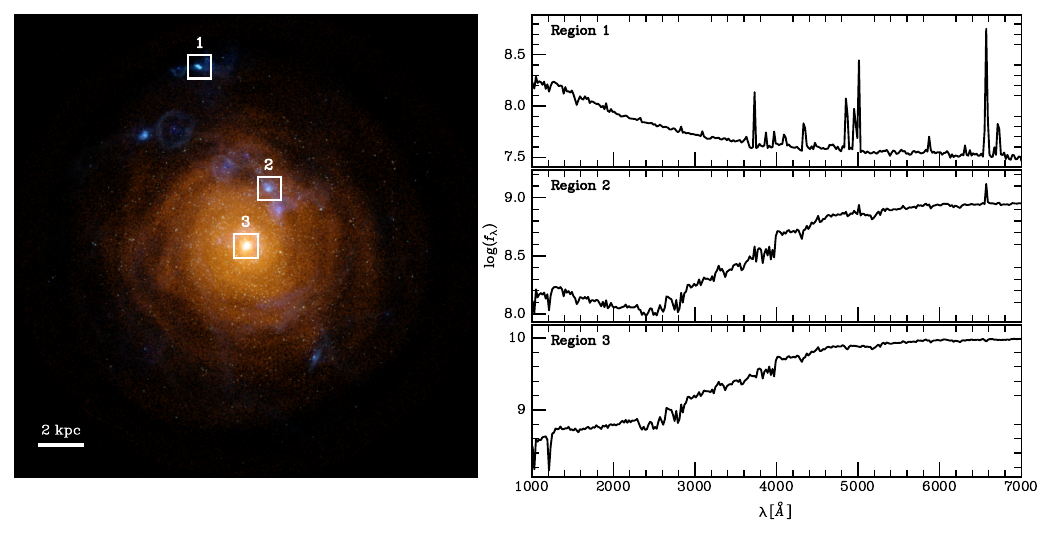}
    \caption{Mock observations of the LMC-like simulation. The left panel shows an RGB image, where red shows the underlying optical continuum (6000--6600\AA), green shows the brightest regions in the near-UV (2300--2600\AA), and blue shows the brightest regions in the far-UV (1200--1500\AA). The pixel scale is 20\,pc and the image is smoothed with a 20\,pc Gaussian kernel. Three 1$\times$1\,kpc boxes are highlighted showing: a star-forming region in the galaxy outskirts (region 1), a star-forming region in the mid-disk (region 2), and the galactic core (region 3). The spectrum of each highlighted region is shown in the right panel in $\log\,f_\lambda$ from 1000-7000\,\AA. \textit{Region 1} shows the characteristic spectrum of a star-forming region, including a steep $\beta$ slope, a Balmer jump, and high-EW emission lines. \textit{Region 2} is the mid-disk, and so the star-forming region is dust-obscured, leading to low-EW line emission and an optical continuum dominated by the old stellar population. \textit{Region 3}, the galactic core, is completely dominated by the old stellar population of bulge stars.}
    \label{fig:lmc_ifu}
\end{figure*}

To demonstrate the mock observables which can be generated with \colt, we create a high-resolution IFS datacube from the LMC-like simulation. We use the converged temperatures and densities as described in the previous section. We perform another pass of the MCRT to model the non-ionizing emission (stellar continuum, nebular continuum, and lines), including absorption and anisotropic scattering by dust. We adopt a Milky Way dust model \citep{Weingartner:2001aa}. 

In Fig.~\ref{fig:lmc_ifu} shows an RGB image, where red corresponds to the underlying optical continuum (6000--6600\AA). Due to the large population of old stars, the optical continuum dominates emission across most of the galaxy, and so we scale the channels such that green highlights the brightest regions in the near-UV (2300--2600\AA), and blue highlights the brightest regions in the far-UV (1200--1500\AA). The pixel scale is 20\,pc and we apply a 20\,pc Gaussian kernel to smooth the image.

We highlight three separate 1$\times$1\,kpc regions in the image and show the corresponding spectrum for each region between 1000-7000\,\AA to illustrate the diverse emergent spectra across a variety of environments. Region 1 is a star-forming complex in the galactic outskirts, and exhibits a steep $\beta$ slope, nebular continuum emission in the form of a Balmer jump, and prominent emission lines. Region 2 also focuses on a star-forming region, but this region is located in the mid-disk and so the ongoing star formation is shrouded in dust. This leads to low equivalent width (EW) emission lines and an optical continuum dominated by the large population of older stars, which are less dust-obscured. Region 3 is centered on the core of the galaxy, and the spectrum here is completely dominated by the large population of old stars which comprise the bulge.

\section{Nebular emission of high-redshift galaxies}
\label{sec:Nebular emission of high-redshift galaxies}

\begin{figure*}
    \centering
	\includegraphics[width=\textwidth]{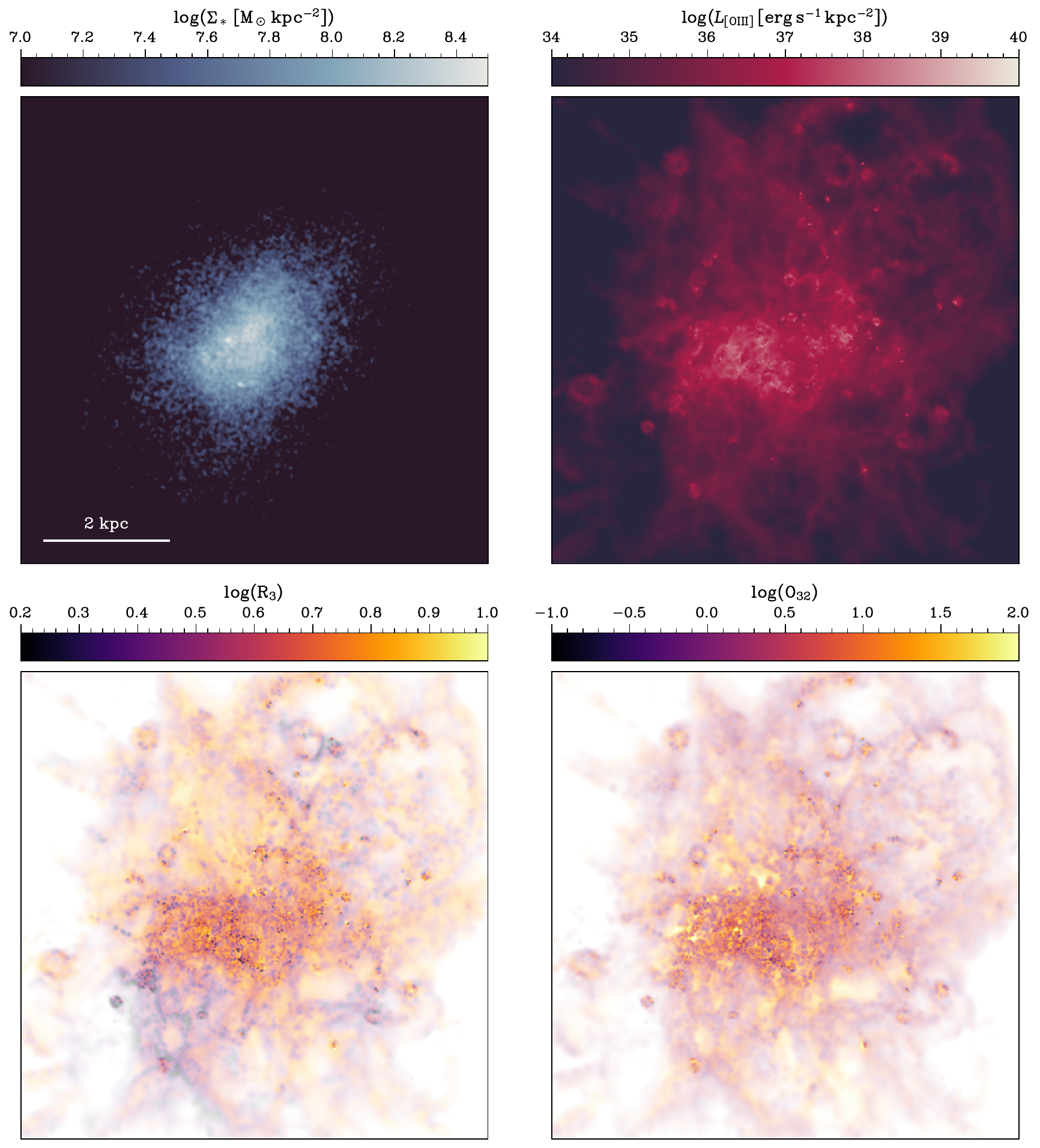}
    \caption{A galaxy from the \thzoom simulations (g205, group 3) at $z=6$ with $M_\ast=10^{8.7}\,\mathrm{M_\odot}$. The images have a pixel-scale of 7.5\,pc. In the upper left panel, we show the stellar mass surface density, smoothed with a 30\,pc Gaussian kernel. In the upper right panel, we show the [\ion{O}{III}]$\lambda$5007~\AA\ surface brightness, smoothed with a Gaussian kernel based on the cell radii. The large-scale, wispy distribution of gas is visible through low surface-brightness emission, whereas the ISM is host to compact, bright structures, which are a tell-tale sign of ongoing star formation. In the lower left panel we show the R23 line ratio ([\ion{O}{III}]$\lambda$5007~\AA\,/ H$\beta$) and in the lower right panel we show the O32 line ratio ([\ion{O}{III}]$\lambda$5007~\AA\,/ [\ion{O}{II}]$\lambda$3727,3730~\AA), with transparency scaled based on the [\ion{O}{III}]$\lambda$5007~\AA\ surface brightness. The line ratios show remarkable spatial variation, both within the ISM and on larger scales.}
    \label{fig:zoom_image}
\end{figure*}

\subsection{Simulation details}
\label{sec:Simulation details}

\begin{figure}
    \centering
	\includegraphics[width=\columnwidth]{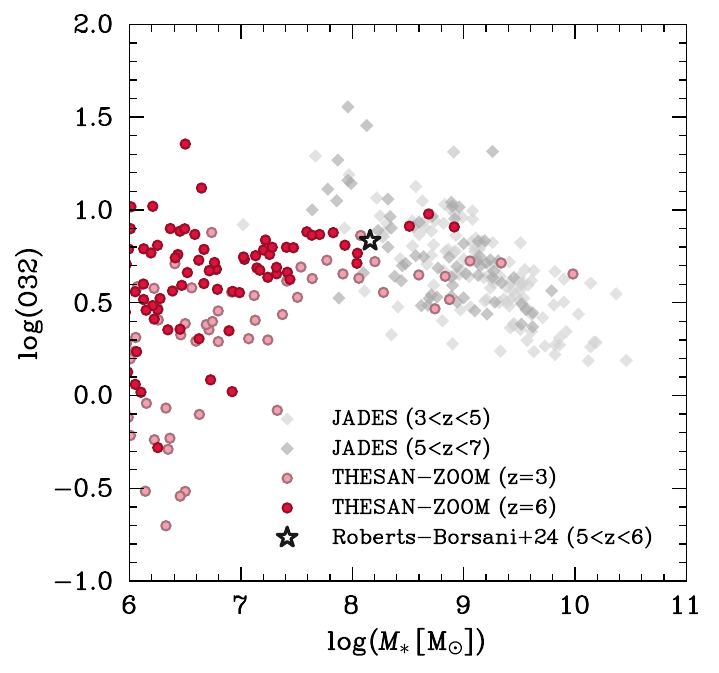}
    \caption{The O32 line ratio ([\ion{O}{III}]$\lambda$5007~\AA\,/ [\ion{O}{II}]$\lambda$3727,3730~\AA) as a function of stellar mass, $M_\ast$. We show galaxies in two snapshots ($z=3$ and $z=6$) of the g205 zoom-in region from the \thzoom simulations, where the line emission has been calculated using \colt. We compare to JADES NIRSpec PRISM measurements split into two redshift bins at $3<z<5$ and $5<z<7$ \citep{DEugenio:2025aa} and the stacked measurements from \citet{Roberts-Borsani:2024aa}. Where observations are available ($M_\ast\approx10^8\,\mathrm{M_\odot}$), the simulations show good agreement, including a decreasing O32 with $M_\ast$. They also both indicate increasing average O32 with redshift at a fixed $M_\ast$. At lower masses, the simulated galaxies plateau in their average O32, however, the scatter is markedly larger. O32 is closely related to ionization parameter and sSFR, and so this scatter may be directly related to the highly stochastic star formation characteristic of these galaxies \citep{McClymont:2025aa}.}
    \label{fig:o32_mstar}
\end{figure}

\begin{figure}
    \centering
	\includegraphics[width=\columnwidth]{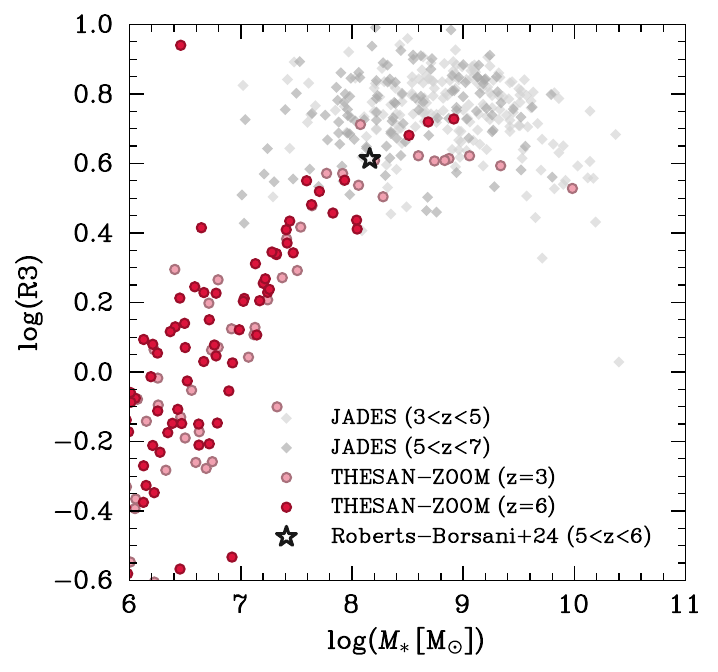}
    \caption{The R3 line ratio ([\ion{O}{III}]$\lambda$5007~\AA\,/ H$\beta$) as a function of $M_\ast$. We show galaxies in two snapshots ($z=3$ and $z=6$) of the g205 zoom-in region from the \thzoom simulations, where the line emission has been calculated using \colt. We compare to JADES NIRSpec PRISM measurements \citep{DEugenio:2025aa} split into two redshift bins at $3<z<5$ and $5<z<7$ and the stacked measurements from \citet{Roberts-Borsani:2024aa}. Both the simulations and observations show a turnover at $M_\ast\approx10^9\,\mathrm{M_\odot}$, which is a well-known effect of increasing metallicity in increasingly massive galaxies. }
    \label{fig:r3_mstar}
\end{figure}

The relative and absolute strength of nebular emission evolves with redshift \citep[e.g.][]{Roberts-Borsani:2024aa}. A key application of the code is to better understand the nature of this evolution, and therefore gain insight into the ISM of galaxies at early cosmic times. In order to demonstrate the proof-of-concept capability for this task, we now apply the code to the \thzoom simulations \citep{Kannan:2025aa}, which are a suite of high-resolution zoom-in simulations targeting galaxies selected from the original large (95.5\,cMpc) \thesan volume \citep{Kannan:2022aa, Smith:2022ab, Garaldi:2022aa,Garaldi:2024aa}. Initial works which have been conducted using \thzoom have studied galaxy-scale star-formation efficiencies \citep[SFEs;][]{Shen:2025aa}, the impact of reionization on galaxies \citep{Zier:2025aa}, Population III star formation \citep{Zier:2025ab}, high-redshift galaxy sizes \citep{McClymont:2025ab}, SFEs on the scale of giant molecular clouds \citep[GMCs;][]{Wang:2025aa}, and chemical evolution \citep{McClymont:2025ae}.

The simulations were carried out with {\sc arepo-rt} \cite{Kannan:2019aa,Zier:2024aa}, a radiation hydrodynamics extension of the moving-mesh code {\sc arepo} \citep{Springel:2010aa,Weinberger:2020aa}. {\sc arepo-rt} performs on-the-fly radiative transfer with a moment-based approach and uses a reduced speed of light approximation to improve computational performance. A primordial non-equilibrium thermochemical network is employed, tracking $\HM, \HI, \HII, \HeI, \HeII,$ and $\HeIII$. Metal cooling is implemented assuming equilibrium with a \citet{Faucher-Giguere:2009aa} UV background. Metal cooling rates are stored in look-up tables \citep{Vogelsberger:2013aa} based on calculations with \textsc{cloudy}. 

\thzoom includes simulations across three resolution levels, which are factors 4, 8, and 16 improvements in the spatial resolution relative to the \thesan simulations (named ``4$\times$'', ``8$\times$'', and ``16$\times$''). The increase in spatial resolution corresponds to an increase of factors of 64, 512, and 4096 in the mass resolution. In this work we will only use snapshots from the 4$\times$ resolution level, which has a baryonic mass resolution of $9.09\times10^3\,\mathrm{M}_\odot$. While there are simulations which are run with physics variations, we exclusively use the fiducial physics runs in this work.

Crucially for this work, the simulations aim to model and resolve the multi-phase ISM, which is in contrast to simulations employing an effective equation-of-state galaxy formation approach \citep[e.g., ][]{Springel:2003aa,Vogelsberger:2013aa,Pillepich:2018aa}, which explicitly do not resolve separate phases in the ISM. \colt takes gas properties, such as density at face-value, so it is most suited to simulations such as \thzoom which faithfully model the ISM. As for the LMC simulation, the \thzoom simulations largely follow the SMUGGLE star formation and feedback model. Stellar feedback includes photoionization, radiation pressure, stellar winds, and supernova feedback \citep{Marinacci:2019aa,Kannan:2020aa,Kannan:2021aa}. Stellar radiation is treated by injecting ionizing radiation, the impact of which is treated self-consistently by the on-the-fly radiative transfer. Supernova feedback injects thermal energy and momentum into the surrounding gas. Stellar winds follow the prescription of the SMUGGLE model described in \citet{Marinacci:2019aa}. An additional early stellar feedback channel is implemented to disrupt molecular clouds shortly after the onset of star formation, which prevents excessive star formation in dense regions. This additional channel is physically motivated by a number of processes which are generally not modeled in large numerical simulations, such as Lyman-$\alpha$ feedback \citep{Smith:2017ab,Nebrin:2025aa}.

\subsection{Nebular line emission}
\label{sec:Nebular line emission high-redshift}

For this proof-of-concept, we run \colt with our standard approach on only two snapshots; as a detailed analysis of the full \thzoom suite will be carried out in a forthcoming paper. We select the g205 zoom-in region (see \citealt{Kannan:2025aa} for details), and choose a snapshot at $z=3$ and $z=6$. As with the LMC simulation, we start with the on-the-fly H and He ionization fractions and run the code to initial convergence with the ionization solver only using $10^7$ photons. We switch the thermal equilibrium solver on and proceed to run again to convergence with $10^7$ photons, and then with $10^8$ photons. We use the Courant-limited cooling option. We do not include a UVB background. Emission line fluxes are intrinsic (no dust) and are calculated within the stellar half-mass radius. In order to exclude emission from gas cells outside of our cooling network, which currently does not include molecular cooling, we exclude gas cells with temperatures below 5000\,K.

Before looking at an analysis of the integrated emission properties, it is first informative to take a spatially resolved view. In Fig.~\ref{fig:zoom_image}, we show images of individual galaxy at $z=6$ (g205, group 3; see \citealt{Kannan:2025aa} for details). The galaxy has a stellar mass of $M_\ast=10^{8.7}\,\mathrm{M_\odot}$ and the pixel scale is 7.5\,pc. The stellar-mass surface density map, which is smoothed with a 30\,pc Gaussian kernal, shows that the bulk of the stellar mass is compact, located in a $\sim1$\,kpc region. The \OIII\ emission line map reveals significantly more extended structures, with the larger-scale gas distribution being visible in low surface-brightness emission. Within the ISM, small-scale ($\lesssim1$\,pixel) bright regions are visible, which represent dense, star-forming complexes. We also show spatially resolved maps of the R3 ([\ion{O}{III}]$\lambda$5007~\AA\,/ H$\beta$) and O32 ([\ion{O}{III}]$\lambda$5007~\AA\,/ [\ion{O}{II}]$\lambda$3727,3730~\AA) line ratios. These emission line ratios show remarkable variation within the galaxy, corresponding to the local conditions, such as the intensity and hardness of the ionizing radiation, the gas-phase metallicity, and the gas density.

We compare the integrated emission properties of the galaxies to emission line ratios of high-redshift galaxies observed with \textit{JWST} NIRSpec as a part of the \textit{JWST} Advanced Deep Extragalactic Survey \citetext{JADES, \citealp{Eisenstein:2023aa}; PID 1180 and~1181, PI: D. Eisenstein; PIDs: 1210 and~1286, PI: N. L\"utzgendorf.  \citealp{Bunker:2023aa,DEugenio:2025aa}}, later extended with parallel NIRSpec observations of the JADES Origins Field \citetext{PID 3215, PIs: D.~Eisenstein and R.~Maiolino; \citealp{Eisenstein:2023ab}}. The emission line fluxes are from the third JADES public release data, presented in \citet{DEugenio:2025aa}. We exclusively use PRISM emission line fluxes in this work and require that all of the lines included in a given emission line ratio have a $\mathrm{S/N}>5$ in order to be plotted. We use stellar masses based on fits first presented in \citet{Simmonds:2024ab}, which are derived with the spectral energy distribution (SED) fitting code \texttt{Prospector} \citep{Johnson:2019aa,Johnson:2021aa}, on the full JADES photometry set. We also show the line ratio for the stack of $5<z<6$ galaxies from \citet{Roberts-Borsani:2024aa}. While still subject to selection biases, this stacked value is less vulnerable to S/N biasing than individual galaxies.

In Fig.~\ref{fig:o32_mstar}, we show the O32 line ratio as a function of stellar mass, $M_\ast$, which is measured within twice the stellar half-mass radius. Compared to galaxies in JADES at $M_\ast>10^8\,\mathrm{M_\odot}$, we find broadly consistent O32 ratios in \thzoom using \colt. Interestingly, we find an apparent turnover/flattening in O32 at $M_\ast\approx10^8\,\mathrm{M_\odot}$. This lower mass regime is generally poorly probed by observations, so it is difficult to assess whether this turnover is seen in observed galaxies. O32 traces the ionization parameter of gas, and is therefore strongly dependent on the specific SFR of the galaxy. This may also explain the large scatter in low-mass galaxies, which have highly stochastic star-formation histories \citep{McClymont:2025aa}. We have not accounted for the observability of galaxies here, and so it is not clear what we would predict the observable trend with mass would be at $M_\ast<10^8\,\mathrm{M_\odot}$. The stacked value from \citet{Roberts-Borsani:2024aa}, which likely represents more typical galaxies than the JADES individual data, lies well within the distribution of \thzoom galaxies. In any case, we can be satisfied that the emission calculated with \colt is capable of reproducing the correct line ratios across a range of stellar masses, and leave more detailed comparison with a larger sample to a future, dedicated study.

In Fig.~\ref{fig:r3_mstar}, we show the R3 line ratio as a function of $M_\ast$. We again see a turnover in this line ratio, although at a somewhat higher mass of $M_\ast\approx10^9\,\mathrm{M_\odot}$. This turnover is a well-known function of metallicity \citep[e.g.,][]{Curti:2020aa}, which is caused by the competing effects on [\ion{O}{III}]$\lambda$5007~\AA\, emissivity with increased metallicity, namely increasing oxygen abundance but decreasing nebular temperature. The simulations appear to reproduce the observations fairly well, although the normalisation may be $\sim$0.1\,dex too low. However, given that the stacked value from \citet{Roberts-Borsani:2024aa} lies close to the simulated galaxies, it is possible that the offset with JADES could be driven by observational biases.

\section{Conclusions}
\label{sec:Conclusions}

We have presented substantial upgrades to the MCRT code \colt, including implementing a thermal equilibrium solver, an atomic level population solver, and nebular continuum emission. The new thermal equilibrium solver accounts for a broad array of physical processes, including photoionization, radiative and dielectronic recombination, collisional ionization, charge exchange, primordial and metal line emission, free-free emission, and Compton scattering of CMB photons. The thermal solver is coupled to the ionization solver and MCRT, allowing us to calculate the ionization states and temperatures of gas self-consistently. The implementation of the thermal equilibrium solver removes a critical limitation with current post-processing methods which rely on temperatures calculated on-the-fly by the simulation. These temperatures inherit the limitations of on-the-fly radiation solvers, such as more limited spectral resolution, less accurate propagation of radiation compared to MCRT post-processing, and increased vulnerability to unresolved Str\"ogren spheres. We also include a cooling limiter based on the sound-crossing time of gas cells, which allows us to accurately calculate the temperature of gas in the ISM, while preserving the on-the-fly temperatures in the CGM and IGM, where the on-the-fly solver is more likely to be accurate.

To facilitate the calculation of the metal line cooling, we implemented an atomic level population solver based on modern atomic data. This level solver also allows us to accurately model a large library of metal emission lines, which is crucial for the comparison of simulations to observations, such as the UV and optical lines made available at high-redshift by \jwst (e.g., \ion{C}{III}]$\lambda$1907,1909~\AA, [\ion{O}{II}]$\lambda$3727,3730~\AA, [\ion{O}{III}]$\lambda$5007~\AA) and the infrared emission lines which can be accessed by \textit{ALMA} (e.g., [\ion{O}{III}]$\lambda$88~$\mu$m, [\ion{C}{II}]$\lambda$158~$\mu$m). We have also implemented primordial nebular continuum emission due to free-free, free-bound, and two-photon emission for both hydrogen and helium, which is particularly relevant given the prevalence of nebular continuum emission in \jwst spectra of high-redshift galaxies \citep[e.g.,][]{Roberts-Borsani:2024aa}.

We validated these new features on several tests, including Str\"omgren spheres and gas in collisional ionization equilibrium, using the \textsc{Cloudy} photoionization code as a benchmark. The agreement between \colt and \textsc{Cloudy} in these tests was excellent, demonstrating the accuracy of our code.

We also applied \colt to a simulation of an LMC-like isolated galaxy, showing that the thermal equilibrium solver reshapes the ISM phase structure where it matters for nebular emission while preserving the diffuse halo. Gas that previously accumulated at lukewarm temperatures ($T=10^3-10^4$\,K) band is largely redistributed into a warm $T\sim10^4$\,K) ionized and a cold ($T<10^3$\,K) neutral phase. These adjustments translate directly into more robust emissivities and line ratios without compromising CGM/IGM structure. We demonstrated the mock observations that can be created with \colt by generating a high-resolution IFS datacube that accounts for the impact of dust, including both absorption and scattering, thereby highlighting the diversity of emergent spectra across the galaxy.

Finally, we showcased the applicability of \colt during the epoch of reionization with a proof-of-concept demonstration using the \thzoom simulations. While a detailed analysis from the entire suite is forthcoming, we generated high-resolution emission line maps, which reveal bright star-forming complexes within the ISM and wispy, low surface brightness emission tracing the larger-scale distribution of gas. We also showed that the R3 and O32 ratios vary strongly across the galaxy, reflecting a dependence on the local radiation and gas properties. We compared integrated R3 and O32 line ratio measurements for snapshots at $z=3$ and $z=6$ to data from \textit{JWST} and found good agreement.

In summary, \colt is now capable of accurately modelling the emission from simulations of galaxies across cosmic time, from the UV to the optical, including contributions from stellar and nebular continuum, as well as nebular line emission. We welcome broad collaboration with the community to further develop \colt as an increasingly useful framework for three-dimensional radiative transfer modelling from star- and galaxy-formation simulations. We currently plan to utilize \colt to investigate a variety of exciting areas, including Population III star formation, purportedly nebular-dominated galaxies and top-heavy IMFs, damped Lyman-$\alpha$ absorption features, the escape of ionizing radiation in UV-bright and line emitting galaxies, and the chemical enrichment of high-redshift galaxies.

\section*{Acknowledgements}

The authors thank Charlotte Simmonds for providing her \texttt{Prospector} fits of JADES galaxies used for the observational comparisons. The authors are grateful to Mark Vogelsberger for access to computing resources on the Engaging cluster at MIT. The authors are grateful to the \thzoom collaboration for facilitating the high-redshift application of \colt in this work. The authors gratefully acknowledge the Gauss Centre for Supercomputing e.V. (\url{www.gauss-centre.eu}) for funding this project by providing computing time on the GCS Supercomputer SuperMUC-NG at Leibniz Supercomputing Centre (\url{www.lrz.de}), under project pn29we. WM thanks the Science and Technology Facilities Council (STFC) Center for Doctoral Training (CDT) in Data Intensive Science at the University of Cambridge (STFC grant number 2742968) for a PhD studentship. WM and ST acknowledge support by the Royal Society Research Grant G125142. AS acknowledges support through HST AR-17859, HST AR-17559, and JWST AR-08709.

Various software packages were used in this work, including \textsc{numpy} \citep{Harris:2020aa}, \textsc{scipy} \citep{Virtanen:2020aa}, \textsc{matplotlib} \citep{Hunter:2007aa}, \textsc{smplotlib} \citep{jiaxuan_li_2023_8126529}, \textsc{swiftascmaps} \citep{borrow_2021_5649259}, and \textsc{astropy} \citep{Astropy-Collaboration:2013aa,Astropy-Collaboration:2018aa,Astropy-Collaboration:2022aa}. 

\section*{Data Availability}

See \href{https://colt.readthedocs.io}{\texttt{colt.readthedocs.io}} for public access and documentation of \colt. All \thzoom simulation data, including snapshots, group, and subhalo catalogs and merger trees, will be made publicly available in the near future. Data will be distributed via \url{www.thesan-project.com}. Before the public data release, data underlying this article will be shared on reasonable request to the corresponding author(s).



\bibliographystyle{mnras}
\bibliography{nebular_colt} 




\appendix




\bsp	
\label{lastpage}
\end{document}